\documentclass[12pt]{article} 

\usepackage{latexsym} 	
\usepackage{amssymb} 
\usepackage{amsbsy}
\usepackage{amsmath}
\usepackage{epsfig}     
\usepackage{graphics,color}
\usepackage{a4}
\usepackage{multirow}
\usepackage{cite}

\usepackage{subcaption}

\numberwithin{equation}{section}

\newcommand{\half}{\frac{1}{2}}

\newcommand{\be}{\begin{equation}}
\newcommand{\ee}{\end{equation}}
\newcommand{\bea}{\begin{eqnarray}}
\newcommand{\eea}{\end{eqnarray}}
\newcommand{\bean}{\begin{eqnarray*}}
\newcommand{\eean}{\end{eqnarray*}}

\newcommand{\hm}{\hspace*{-0.6cm}}

\newcommand{\bit}{\begin{itemize}}
\newcommand{\eit}{\end{itemize}}

\newcommand{\om}{\omega}

\newcommand{\pv}{\mathbf p}

\newcommand{\el}{\vspace*{0.5cm}}

\DeclareMathOperator*{\AICc}{AICc}

\begin{document}

\title{\bf\large
Open charm mesons at nonzero temperature: results in the hadronic phase from lattice QCD
}

\author{
\small
 Gert Aarts$^{a,b}$,
 Chris Allton$^{a}$, 
 Ryan Bignell$^{a}$,
 Timothy J.\ Burns$^{a}$, \\
\small
 Sergio Chaves Garc\'ia-Mascaraque$^{a}$, 
 Simon Hands$^{c}$,
 Benjamin J\"ager$^{d}$, \\
 \small 
 Seyong Kim$^{e,f}$, 
 Sin\'ead M.\ Ryan$^{g}$,
 Jon-Ivar Skullerud$^{g,h}$\thanks{\scriptsize email:  \{g.aarts, c.allton, ryan.bignell, t.burns, 989336\}@swansea.ac.uk, simon.hands@liverpool.ac.uk, jaeger@imada.sdu.dk, skim@sejong.ac.kr, ryan@maths.tcd.ie, jonivar@thphys.nuim.ie}
 \\
\mbox{} \\
{\small 
${}^a$ Department of Physics, Swansea University, Swansea, SA2 8PP,  United Kingdom} \\
{\small 
${}^b$  European Centre for Theoretical Studies in Nuclear Physics and Related Areas  }\\
{\small  
(ECT*) \& Fondazione Bruno Kessler, 
38123 Villazzano (TN), Italy} \\
{\small 
${}^c$ Department of Mathematical Sciences, University of Liverpool}\\ 
{\small 
Liverpool L69 3BX, United Kingdom}\\
{\small 
${}^c$ CP3-Origins \& Danish IAS, Department of Mathematics and Computer Science} \\
{\small 
University of Southern Denmark, 5230, Odense M, Denmark}\\
{\small 
${}^e$ Department of Physics, Sejong University, Seoul 143-747, Korea}\\
{\small 
${}^f$ AEC, Institute for Theoretical Physics, University of Bern, 
Bern, Switzerland}\\
{\small 
${}^g$ School of Mathematics, Trinity College, Dublin, Ireland}\\
{\small 
${}^h$ Department of Theoretical Physics, National University of Ireland Maynooth, }\\
{\small  
County Kildare, Ireland}\\
}

\date{\normalsize September 27, 2022}

\maketitle

\vspace*{-0.5cm}

\begin{abstract}
{
We study what happens to $D$ and $D_s$ mesons as the temperature increases, using lattice QCD simulations with $N_f=2+1$ dynamical  flavours on anistropic lattices. 
We have access to five temperatures in the hadronic phase. 
Using the determined groundstate mass at the lowest temperature, we investigate the effect of rising temperature by analysing ratios of mesonic correlators, without the need for further fitting or spectral reconstruction.  
In the pseudoscalar and vector channels, we demonstrate that temperature effects are at the percent level and can be captured by a reduction of the groundstate mass as the thermal crossover is approached. 
In the axial-vector and scalar channels on the other hand, temperature effects are prominent throughout the hadronic phase. 
}
\end{abstract}

\maketitle


 
\tableofcontents

\section{Introduction}
 \label{sec:intro}

The fate of hadrons under extreme conditions is one of the outstanding questions in the theory of strong interactions, Quantum Chromodynamics (QCD). As the temperature increases, the hadron gas---with confined quarks and broken chiral symmetry---smoothly \cite{Aoki:2006we} transitions into a quark-gluon plasma (QGP), with deconfined light degrees of freedom and chiral symmetry restored. Clear evidence for this comes from nonperturbative simulations of lattice QCD, analysing for instance the behaviour of the pressure, entropy and quark number susceptibilities across the transition, as well as the chiral condensate and its susceptibility 
\cite{Borsanyi:2010bp,Bazavov:2011nk,Borsanyi:2010cj,HotQCD:2014kol,Borsanyi:2011sw,HotQCD:2012fhj}.
For hadrons built out of heavier (charm and bottom) quarks, the survival of bound states in the QGP is possible and e.g.\ the melting pattern of bottomonium states provides important insight into the length scales in the QGP \cite{Burnier:2007qm,Brambilla:2010vq,Strickland:2011aa}, which can be studied using lattice QCD \cite{Aarts:2011sm,Aarts:2014cda,Kim:2014iga,Larsen:2019zqv}.

The behaviour of light and strange baryons at nonzero temperature has been studied in detail in Refs.\  \cite{Aarts:2015mma,Aarts:2017rrl,Aarts:2018glk}. The focus there was on parity doubling, a signal for chiral symmetry restoration. As the transition is a crossover, and not a proper phase transition, one may expect a precursor to chiral symmetry restoration already in the hadronic phase, and this was indeed observed, via a reduction of the groundstate masses for baryons with negative parity.  A near-degeneracy of baryons with positive and negative parity at the crossover is consistent with the notion of chiral symmetry being restored. 

In this paper we consider heavy-light systems: mesons built out of a charm quark and either a light or strange quark, i.e.\ $D$ and $D_s$ mesons. Charm quarks have been of interest to QGP phenomenology since the dawn of the field, with $J/\psi$ suppression as one of the proposed signatures of the formation of the QGP \cite{Matsui:1986dk}. 
At high temperature, equilibration and thermalisation of charm quarks yield insight into the transport properties of the QGP \cite{Moore:2004tg,vanHees:2004gq}, while at lower temperature 
the formation of open charm states, in particular $D$ mesons,  through recombination or coalesence, gives information on charm-quark interactions in the thermal medium \cite{Greco:2003vf}.  The propagation of $D$ mesons in the hadronic phase and possible consequences for heavy-flavour observables have been investigated in Ref.\ \cite{Ozvenchuk:2014rpa}. Additional references can be found in e.g.\ Refs.\ \cite{Aarts:2016hap,Zhao:2020jqu,Beraudo:2022dpz}.

In this work we are interested in the response of $D_{(s)}$ mesons to an increasing temperature, i.e.\ starting at low temperature in the hadronic phase. Hence a natural point of comparison is not perturbative QCD, but instead effective hadronic theories. In a series of papers, this topic was addressed recently by Monta\~na et al \cite{Montana:2020lfi,Montana:2020vjg,Montana:2020var,Montana:2022bxe} and we will compare our findings to the results presented there. Other studies of thermal effects on $D_{(s)}$ mesons using effective models include Refs.\ \cite{Fuchs:2004fh,Sasaki:2014asa,Buchheim:2018kss}.
Indeed, one motivation for our work is to provide a first-principle benchmark for effective descriptions that aim to describe hadrons under extreme conditions, notably since those models can be extended to regions of the phase diagram where Monte Carlo simulations of lattice QCD are not directly applicable, such as at larger baryon chemical potential. 
Previous studies of open charm using lattice QCD include an analysis of cumulants of net charm fluctuations \cite{Bazavov:2014yba}, of screening masses in the $D_s$ meson channel in the QGP \cite{Bazavov:2014cta}, and of spectral functions obtained from $D$ and $D_s$  meson correlators on anisotropic lattices \cite{Kelly:2018hsi}. The data of the latter is compared to effective hadronic models in Ref.\ \cite{Montana:2020var}.

This paper is organised as follows. In Sec.\ \ref{sec:idea} we discuss the (well-known) difficulty in extracting spectral information from Euclidean lattice correlators and propose a simple way to analyse temperature dependence of correlators without the need for explicit spectral reconstruction. This method is further analysed in Sec.~\ref{sec:model}, using a simple model. We then apply this approach first to $D_{(s)}$ correlators in the 
pseudoscalar ($D, D_s$) and vector ($D^*, D^*_s$) channels in Sec.~\ref{sec:PSV} and subsequently to scalar ($D^*_{0}, D^*_{s0}$) and axial-vector ($D_{1}, D_{s1}$) channels in Sec.~\ref{sec:SAX}. Conclusions are drawn in Sec.~\ref{sec:conclusion}. 
Aspects of the finite-temperature lattice ensembles are summarised in App.\ \ref{sec:lattice}, while App.\ \ref{sec:G-rec} contains a brief comparison with so-called reconstructed correlators. App.\ \ref{sec:fit} finally contains details of the method used to extract the groundstate mass.

\section{Comparing thermal correlators}
\label{sec:idea}


The problem of extracting spectral information from simulations on a Euclidean lattice is well known \cite{Asakawa:2000tr,Aarts:2002cc,Meyer:2011gj,Rothkopf:2019ipj}: the desired information is most clearly visible in spectral functions $\rho(\om,\pv;T)$, but one only has access to numerically determined correlators $G(\tau, \pv;T)$. These two quantities are related via the integral relation (see e.g.\ Ref.\ \cite{Laine:2016hma})
\be
\label{eq:Grho}
G(\tau, \pv;T) = \int_0^\infty \frac{d\om}{2\pi}\, K(\tau,\om;T) \rho(\om, \pv;T), 
\ee
where for mesonic correlators the kernel reads (generalised Laplace transform, $0\leq \tau < 1/T$)
\be
\label{eq:K}
K(\tau, \om;T) = \frac{\cosh[\om(\tau-1/2T)]}{\sinh(\om/2T)} = e^{-\om \tau} [1+n_B(\om)] + e^{\om \tau} n_B(\om),
\ee
with $n_B(\om)=1/[\exp(\om/T)-1]$ the Bose distribution.
A direct inversion of relation (\ref{eq:Grho}) is a classic ill-posed problem \cite{Asakawa:2000tr}.
From now on we work at vanishing momentum and drop the $\pv$ label. 

\begin{figure}[t]
\begin{center}
\includegraphics[width=0.75\textwidth]{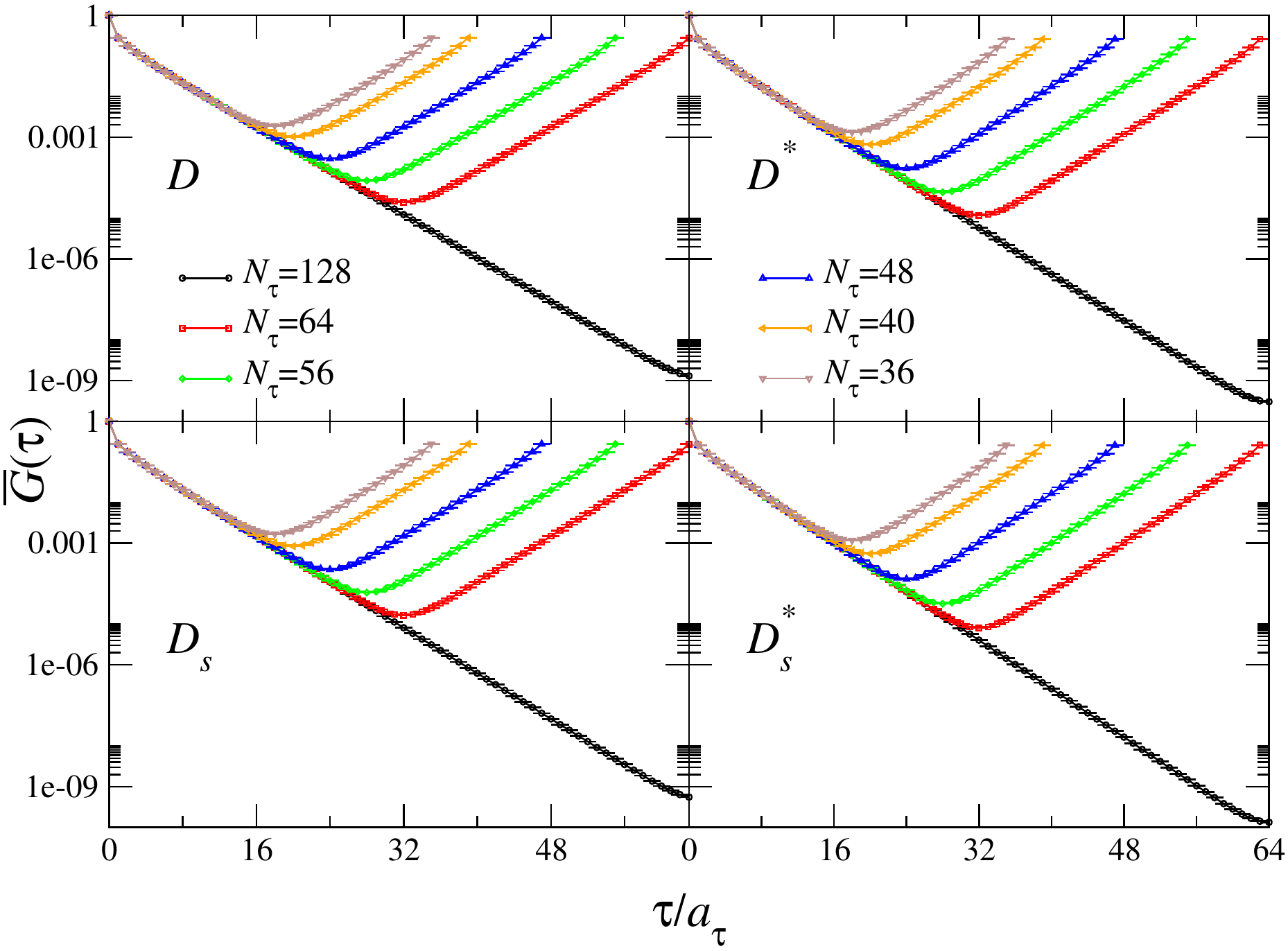}
\end{center}
  \caption{$D_{(s)}$ and $D_{(s)}^*$ correlators $\overline G(\tau)= G(\tau)/G(0)$ at six temporal lattice extents $N_\tau$, or six temperatures $T=1/a_\tau N_\tau$. The correlators on the $N_\tau=128$ lattice are shown up to $\tau/a_\tau=64$ for clarity. Errors are included but are smaller than the symbols.
    Details of these correlators will be discussed below.
  }  
    \label{fig:G-D-Ds-PS-V}
\end{figure}

\begin{figure}[t]
\begin{center}
\includegraphics[width=0.75\textwidth]{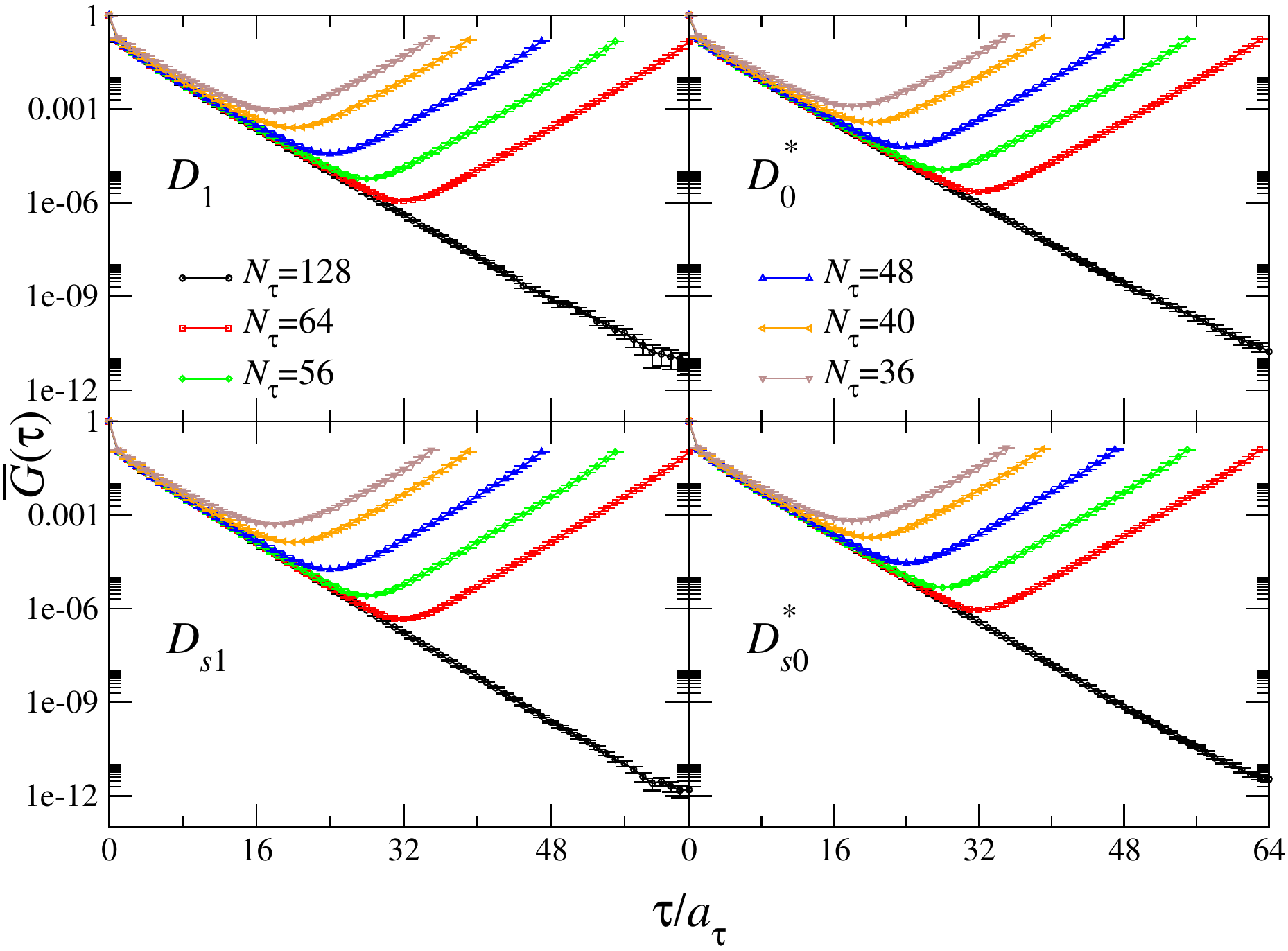} 
\end{center}
  \caption{As in Fig.\ \ref{fig:G-D-Ds-PS-V}, for the $D_{1(s)}$ and $D_{0(s)}^*$ correlators.} 
    \label{fig:G-D-Ds-AV-SC}
\end{figure}

A feature of Eq.\ (\ref{eq:Grho}) is that a correlator will always show temperature dependence when comparing different temperatures, even when the physical information encoded in the spectral function is unchanged. The reason is that the kernel is (trivially) temperature dependent, due to the way temperature is encoded in the compact Euclidean direction. On a lattice with temporal lattice spacing $a_\tau$ and number of timeslices $N_\tau$, the temperature is given by $T=1/a_\tau N_\tau$. In a fixed-scale approach, which we follow here, higher temperature corresponds to smaller $N_\tau$, and correlators will deviate from the ones at lower temperatures at earlier and earlier Euclidean times, as illustrated in Figs.\ \ref{fig:G-D-Ds-PS-V} and \ref{fig:G-D-Ds-AV-SC}, irrespective of whether spectral information is changing.\footnote{Details of the correlators and lattice ensembles are discussed further below and in App.~\ref{sec:lattice}. 
Note that throughout the paper we always consider normalised correlators, obtained as 
\be
\label{eq:Gnorm}
\overline G(\tau;T)  = G(\tau;T)/G(0;T),
\ee
but to avoid a cluttering of symbols we will not indicate the overline in this section.}

 One attempt to eliminate this temperature effect is to use so-called {\em reconstructed correlators} \cite{Datta:2003ww}: one constructs a spectral function at a reference (typically the lowest) temperature---which we will denote as $T=T_0$---only and subsequently integrates this spectral function with the $T$-dependent kernel (\ref{eq:K}). This yields a reconstructed correlator at temperature $T$ under the assumption that the physical information has not changed from the reference temperature $T_0$,
\be
G_{\rm recon}(\tau;T,T_0) = \int_0^\infty \frac{d\om}{2\pi}\, K(\tau,\om;T) \rho(\om;T_0).
\ee
 Comparing the reconstructed and the actual correlator at temperature $T$, via the ratio
 \be
 R_{\rm recon}(\tau; T, T_0) = G(\tau;T)/G_{\rm recon}(\tau;T,T_0),
 \ee
allows one to assess their similarity. In the case that they differ, one concludes that the actual spectral function at temperature $T$, and therefore the physical content, has changed.

This approach still requires the construction of a spectral function at the reference temperature $T_0$.\footnote{In the special case that the ratio of temperatures is an integer, this can be avoided, see Ref.\ \cite{Ding:2012sp} and App.\ \ref{sec:G-rec}.}
 One may wonder, however, whether it is necessary to construct the full spectral function at all energies, in particular since spectral functions from lattice correlators contain physically irrelevant  information at higher energies, which is not of practical interest \cite{Aarts:2005hg}. 
  The most minimal approach is to extract only the mass of the groundstate at $T=T_0$, provided it is well defined. This implies fitting\footnote{The word `fitting' is used here as a shorthand for the method detailed in App.~\ref{sec:fit} or any other approach to extract the groundstate mass.} the exponential decay of the correlator, made periodic at $T\not=0$, as seen in  Figs.\ \ref{fig:G-D-Ds-PS-V} and \ref{fig:G-D-Ds-AV-SC}. In terms of spectral functions and correlators, this amounts to the Ansatz,
 \bea
 \rho_{\rm single\, peak}(\om;T) &&\hm= A(T) 2\pi\delta\left[\om-M(T)\right], \\
 \label{eq:peak}
G_{\rm single\, peak}(\tau;T) &&\hm= A(T) \frac{\cosh[M(T)(\tau-1/2T)]}{\sinh[M(T)/2T]},
 \eea
where both the mass $M(T)$ and amplitude $A(T)$ may depend on temperature. The assumption is that any possible width of the groundstate peak in the spectral function is negligible.
Note that in the correlator the groundstate is only clearly visible at larger Euclidean times,\footnote{Due to the periodicity, `large' is always understood as $0\ll \tau \le 1/2T$.} i.e.\ once the effect of excited states are suppressed.  
As the temperature increases, continuing with a fitting approach leads to a similar problem as before: the temporal extent of the lattice is reduced and the groundstate may no longer be disentangled from excited states, even in the case that the spectral information is unchanged. 
In addition, periodicity leads to a bend in the correlator around $\tau=1/2T$, or $\tau/a_\tau=N_\tau/2$, obscuring the decay, see again Figs.\ \ref{fig:G-D-Ds-PS-V} and \ref{fig:G-D-Ds-AV-SC}.

One can eliminate both effects by dividing the correlator at temperature $T$ with a model correlator, 
\be
\label{eq:Gmodel}
G_{\rm model}(\tau;T, T_0) = A(T_0) \frac{\cosh[M(T_0)(\tau-1/2T)]}{\sinh[M(T_0)/2T]}.
\ee
in which the parameters determined at the reference temperature $T_0$ are used, and consider the ratio
\be
\label{eq:r}
r(\tau;T, T_0) = G(\tau;T)/G_{\rm model}(\tau; T, T_0).
\ee
Provided that the groundstate properties are precisely determined, this ratio will be independent of $\tau$ for larger Euclidean time, in the region where the groundstate dominates. It is  however, not completely $\tau$ independent, as e.g.\ excited states are present in the numerator but not in the denominator. 

Therefore, as a final step and motivated by the concept of the reconstructed correlator, we propose to take the ratio of the combination above at  temperatures $T$ and $T_0$, and consider
\be
\label{eq:R}
R(\tau; T, T_0) = \frac{G(\tau;T)}{G_{\rm model}(\tau; T, T_0)} \Bigg/ \frac{G(\tau; T_0)}{G_{\rm model}(\tau;T_0, T_0)}.
\ee 
Note that this double ratio can also be written as
\be
\label{eq:R2}
R(\tau; T, T_0) = \frac{G(\tau; T)}{G(\tau; T_0)} \Bigg/ \frac{G_{\rm model}(\tau; T, T_0)}{G_{\rm model}(\tau; T_0, T_0)},
\ee 
i.e.\ it is the ratio of correlators and model correlators. 
One can view this double ratio as a {\em poor man's reconstructed correlator}: rather than using the full spectral function at the reference temperature $T_0$, we only use the information on the groundstate at $T=T_0$, under the assumption that it is well described by a narrow peak. As mentioned, this information can be extracted (and a posteriori justified) using standard fits. As above, deviations from 1 imply changes in spectral content. 
These ideas are further tested in the case of a simple model in the next section and applied to $D_{(s)}$ meson lattice data in the remainder of the paper.

\section{Simple model}
\label{sec:model}

To further motivate the analysis of ratios of correlators described above, we consider the following simple model. Let us assume that the  
correlator can be written as the sum of two terms,
\be
\label{eq:G-Ansatz}
G(\tau;T) = A_0 g(\tau; m_0,T) + A_1 g(\tau; m_1,T),
\ee
where 
\be
\label{eq:g}
 g(\tau; m,T) = \frac{\cosh[m(\tau-1/2T)]}{\sinh(m/2T)}
\ee  
is a single-state correlator. $A_{0,1}$ are the amplitudes. Note that this correlator corresponds to a spectral function which is a sum of delta-functions, 
 \be
 \rho(\om) = A_0 2\pi\delta(\om-m_0) +  A_1 2\pi\delta(\om-m_1).
\ee
We interpret this simple Ansatz as follows: $m_0$ is the mass of the groundstate we are interested in. It may or may not depend on the temperature. $m_1$ represents anything beyond the groundstate, e.g.\ a first excited state, but it can be thought of as describing deviations from a single exponential more generally. Of course it is possible to include in Eq.\ (\ref{eq:G-Ansatz}) more terms or terms not of the form (\ref{eq:g}).

The correlator (\ref{eq:G-Ansatz}) may depend on the temperature in two ways: via the explicit $T$ dependence in Eq.\ (\ref{eq:g}), and via the temperature dependence of the masses $m_{0,1}(T)$ and amplitudes  $A_{0,1}(T)$, i.e.\ of the spectral function. Here we are interested in the possible temperature dependence of $m_0$.

As in the preceding section, we compare the normalised correlator,
\be
\label{eq:Gnorm2}
\overline G(\tau;T)  = G(\tau;T)/G(0;T),
\ee
with the model correlator, 
\be
G_{\rm model}(\tau; m_{\rm fit}, T) = g(\tau; m_{\rm fit}, T),
\ee
where $m_{\rm fit}$ is the mass determined via a fitting procedure at a reference temperature $T_0$, and for simplicity we have taken $A_{\rm fit}=1$.
Note that the normalised correlator (\ref{eq:Gnorm2}) using Eq.\ (\ref{eq:G-Ansatz}) depends on $A_1/A_0$; again for simplicity we put $A_1/A_0=1$ as well.
To judge the accuracy of $m_{\rm fit}$, we finally construct the ratio
\be
\frac{\overline G(\tau;T)}{G_{\rm model}(\tau; m_{\rm fit}, T)} = \frac{1}{N} \frac{g(\tau; m_0, T) + g(\tau; m_1, T)}{g(\tau; m_{\rm fit}, T)},
\ee
with $N= g(0; m_0, T) + g(0; m_1, T)$.

\begin{figure}[t]
\begin{center}
\includegraphics[width=0.45\textwidth]{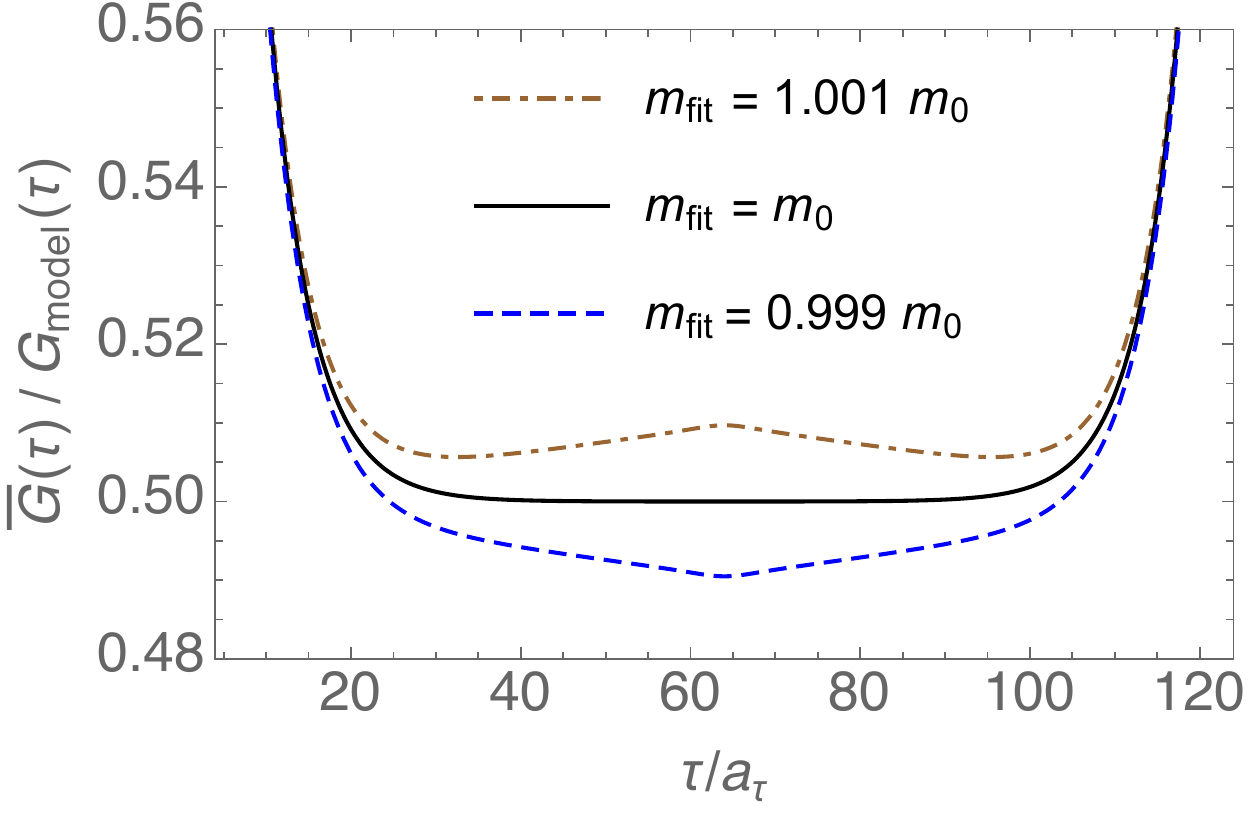}
\includegraphics[width=0.45\textwidth]{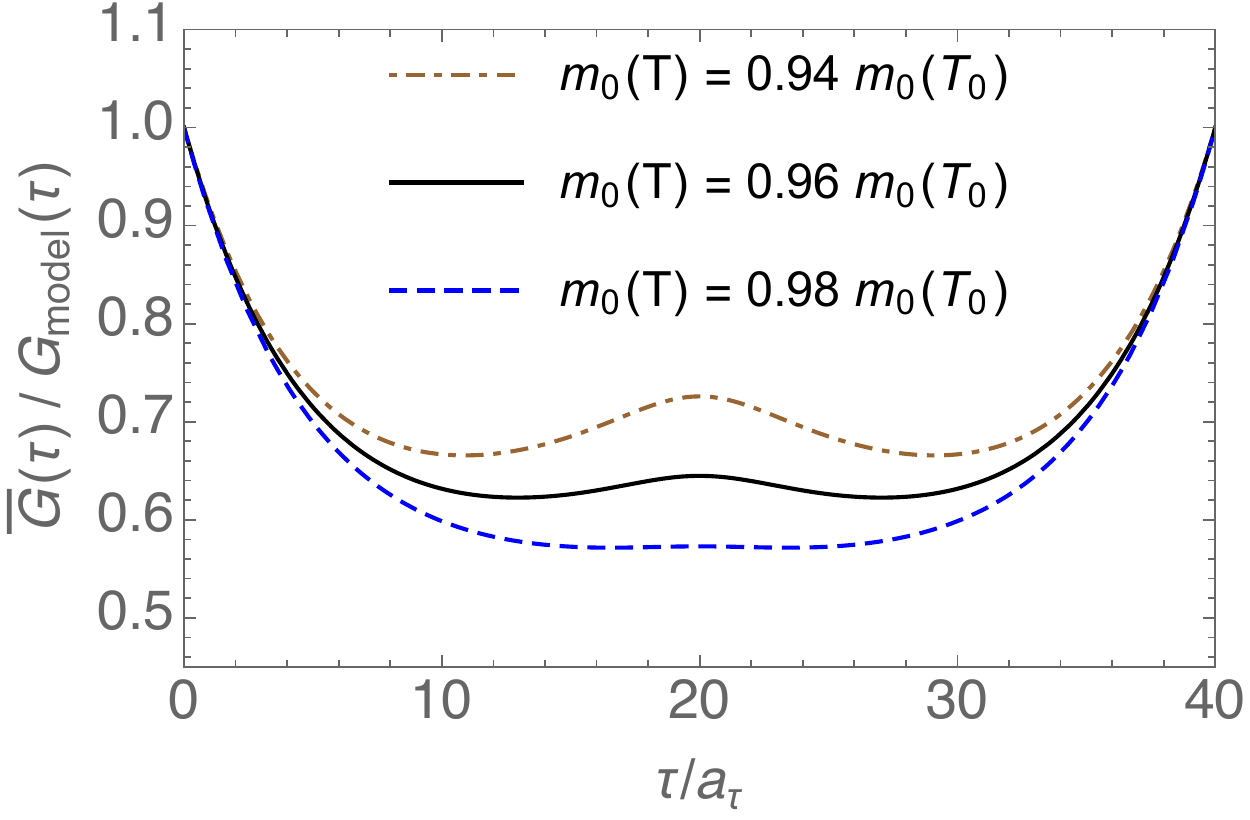}
\end{center}
  \caption{Ratio of the normalised correlator, with groundstate (excited-state) mass $m_0$ ($m_1$), and the model correlator, with mass $m_{\rm fit}$, in the simple model. Left: to investigate the accuracy of the model, $m_{\rm fit}$ is varied around $m_0$ on a lattice with $N_\tau=128$ sites. Right: to investigate a temperature-dependent groundstate mass, $m_0(T)$ is reduced from $m_0(T_0)$, while $m_{\rm fit}$ is kept at $m_0(T_0)$, on a lattice with $N_\tau=40$ sites.
  }
    \label{fig:G-toy}
\end{figure}

We now consider a numerical example motivated by the lattice parameters. Given that the inverse lattice spacing $1/a_\tau\simeq 6$ GeV and $D$ meson masses are around 2 GeV, we take for the groundstate mass $a_\tau m_0=0.3$ (see also Table \ref{tab:masses-at}) and $a_\tau m_1=0.5$. 
In the ideal case, the fitted mass $m_{\rm fit}$ is equal to the groundstate mass $m_0$. Indeed, as illustrated in Fig.\ \ref{fig:G-toy} (left), 
the ratio becomes constant around the centre of the lattice, where excited-state effects are suppressed. If the fitted mass is slightly too large ($m_{\rm fit}=1.001 m_0$) or too small ($m_{\rm fit}=0.999 m_0$), one sees a distinct deviation. Note that the change in mass (1 per mille, $a_\tau m=0.0003$)  corresponds to 2 MeV, an indication of the sensitivity. 
 
Fig.\ \ref{fig:G-toy} (right) illustrates the effect of a decreasing groundstate mass at higher temperature, $m_0(T) < m_0(T_0)$, while the mass in the model correlator $m_{\rm fit}$ is unchanged at $m_0(T_0)$, the value determined at the lowest temperature. A characteristic wiggle around the centre of the lattice is seen in this case, which increases in amplitude for stronger mass reductions.

\section{Pseudoscalar and vector channels}
\label{sec:PSV}

In the remainder of this paper, we apply the concepts developed above to lattice QCD data for $D$ and $D_s$ meson correlators. 
Details of the anisotropic lattice ensembles, with $N_f=2+1$ flavours of dynamical quarks and anisotropy $a_s/a_\tau=3.453(6)$, are given in App.\ \ref{sec:lattice}. Here we summarise the main points. The strange quark is tuned to its physical value, but the light quarks are heavier than in nature. This is quantified by a pion with mass $m_\pi=239(1)$ MeV. We use a fixed-scale approach, in which the lattice spacing is kept fixed, with inverse temporal lattice spacing  $a_\tau^{-1} =6.079(13)$ GeV \cite{Wilson:2019wfr}, and the temperature is increased by decreasing the number of points $N_\tau$ in the temporal direction, using the relation $T=1/a_\tau N_\tau$. The pseudocritical temperature $T_{\rm pc}=167(2)(1)$ MeV is determined via the inflection point of the renormalised chiral condensate \cite{Aarts:2020vyb}.
  We have access to five ensembles in the hadronic phase, as detailed in Table \ref{tab:summary}.

\begin{table}[h]
  \begin{center}
         \begin{tabular}[t]{|c|c|c|c|c|c||c|c|c|c|c|}
	\hline
    	$N_\tau$  		& 128   & 64 	& 56     & 48     & 40     & 36    & 32     & 28     & 24     & 20 \\
	$T$ [MeV]  	&   47   & 95 	& 109   & 127   & 152   & 169  & 190   & 217   & 253   & 304 \\
	\hline
   \end{tabular}
    \end{center}
     \caption{Temporal lattice extents and corresponding temperatures.
     }
          \label{tab:summary} 
\end{table}

\subsection{Groundstate masses at the lowest temperature}

We start with the analysis of the correlators, shown in Figs.\ \ref{fig:G-D-Ds-PS-V} and \ref{fig:G-D-Ds-AV-SC}, at the lowest temperature.  Following the fitting procedure described in App.~\ref{sec:fit}, we obtain the masses given in Table \ref{tab:masses-w-error-T0}. The error in the final column reflects the combined statistical and systematic uncertainty from the fitting procedure as well as the uncertainty from the scale setting.  The former is described in detail in App.~\ref{sec:fit}; the latter is on the order of 0.2\% or about 4 MeV. In the pseudoscalar and vector channels, the error is dominated by the scale setting, in the scalar and axial-vector channels (except in the $D_{s1}$ case) by the fitting procedure; we come back to this below.

 \begin{table}[t]
  \begin{center}
         \begin{tabular}[t]{|c| l |c|c|c|c|}
         \hline
	 		&  				& $J^P$ & PDG [MeV]  	& $a_\tau M$ 	& $M$ [MeV]  \\
	\hline
    	$D$  	& pseudoscalar		& $0^-$ & 1869.65(5) 	& 0.3086(1) 	& 1876(4)  \\
    	$D^{*}$  	& vector			& $1^-$ &  2010.26(5) 	& 0.3291(1)  	& 2001(4) \\
	$D_0^*$	& scalar			& $0^+$ & 2300(19) 		& 0.3656(14) 	& 2222(10) \\
  	$D_1$  	& axial-vector		&$1^+$ &  2420.8(5) 	& 0.3823(70) 	& 2325(43) \\
	\hline
	$D_s$  	& pseudoscalar		& $0^-$ & 1968.34(7)  	& 0.3243(3) 	& 1972(5) \\
	$D_s^*$  	& vector			& $1^-$ &  2112.2(4) 	& 0.3442(1)	& 2092(4) \\
	$D_{s0}^*$ & scalar			& $0^+$ & 2317.8(5) 	& 0.3479(46) 	&  2115(29) \\
 	$D_{s1}$ 	& axial-vector		& $1^+$ & 2459.5(6) 	& 0.4132(2)	& 2512(6) \\
	\hline
   \end{tabular}
    \end{center}
     \caption{$D$ and $D_s$ mesons: PDG mass values \cite{ParticleDataGroup:2020ssz} and our lattice QCD results at $T=47$ MeV.
      The error in the final column reflects the combined statistical and systematic uncertainty from the fitting procedure as well as the uncertainty from the scale setting.      }
     \label{tab:masses-w-error-T0} 
\end{table}

\begin{figure}[t]
\begin{center}
\includegraphics[width=0.7\textwidth]{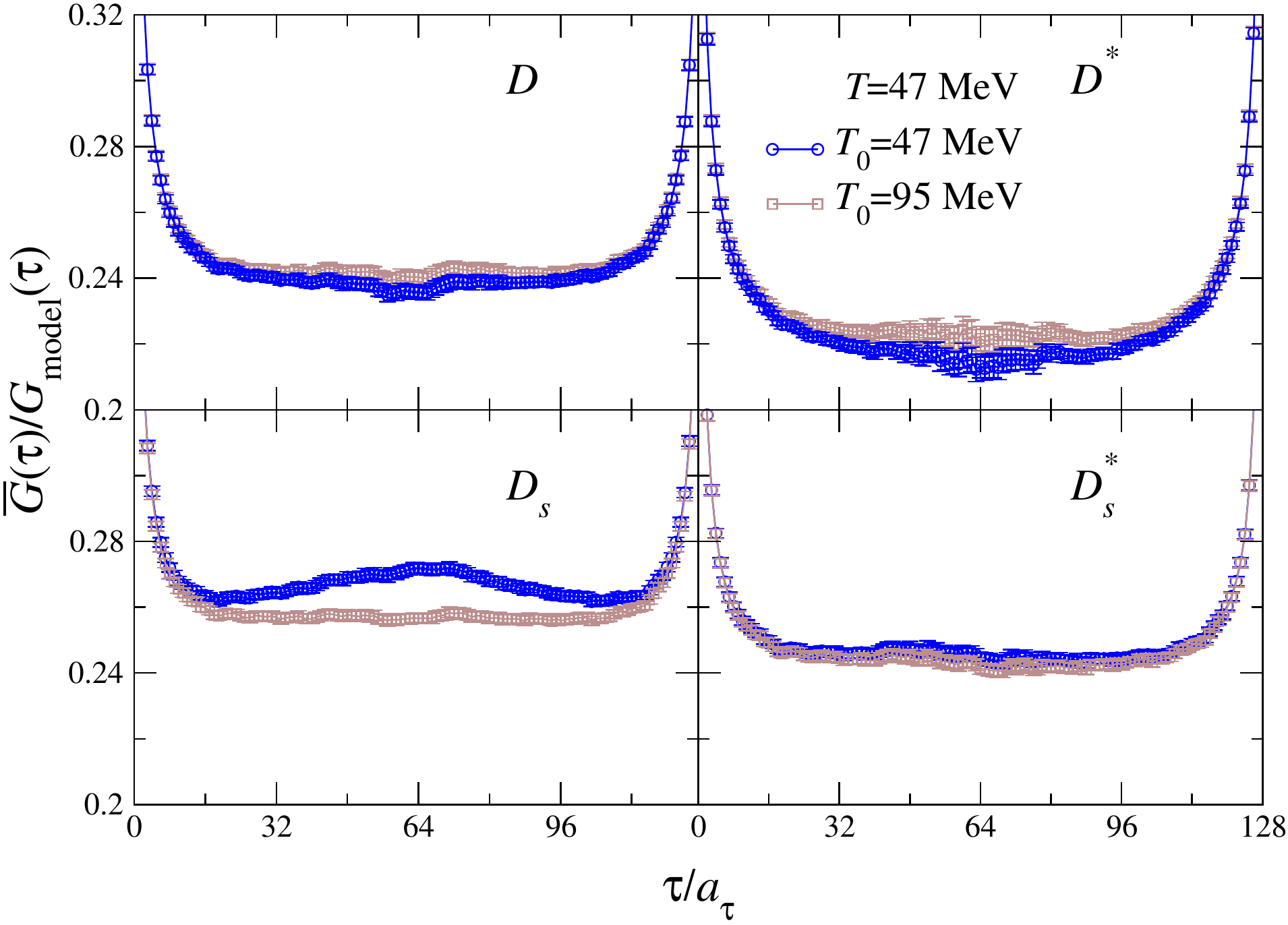} 
\end{center}
  \caption{Ratio $r(\tau;T,T_0)$ of correlator $\overline G(\tau) = G(\tau)/G(0)$ and model correlator $G_{\rm model}(\tau)$ for $D_{(s)}$ and  $D_{(s)}^*$ mesons, at the lowest  temperature, $T=47$ MeV, using  as input the groundstate masses determined at $T_0=47$ MeV (blue circles) and at $T_0=95$ MeV (brown squares). 
  }
    \label{fig:G-ratio-model-T0}
\end{figure}

We continue with a discussion of the pseudoscalar and vector channels; the scalar and axial-vector channels are further analysed in Sec.\ \ref{sec:SAX}. To validate these results, we carry out the analysis described in Sec.\ \ref{sec:idea} and divide the correlator at the lowest temperature with the model correlator, a simple single-state propagator function, see Eq.\ (\ref{eq:r}).
For simplicity we put the amplitude $A(T)=1$; this only affects the vertical scale, but not the shape of the ratio. 
 Using the fitted mass at $T=47$ MeV gives the ratios shown in Fig.\ \ref{fig:G-ratio-model-T0} (blue circles).
Overall, a horizontal plateau indicates that the groundstate is well described by a single exponential. 
Note that this is quite similar to an effective mass plot, as it illustrates at which Euclidean times excited states and other high-energy features cannot be ignored, but also takes the periodicity  in $1/T$ into account.
  
The ratio is very sensitive to the closeness of the fitted mass to the actual value, on the order of a few MeV, see Sec.~\ref{sec:model} and in particular Fig.~\ref{fig:G-toy} (left). This is particularly visible for the $D_s$ meson, for which the ratio is not quite flat  but goes upwards towards the middle of the lattice. This is an indication that the fitted mass is slightly too high (see Sec.\ \ref{sec:model}). Therefore we also present the ratio using the fitted masses at the second-lowest temperature, $T=95$ MeV (brown squares, see Table \ref{tab:masses-at}). 
For the $D_s$ meson, it seems that the fitted value at  $T=95$ MeV gives a better estimate of the groundstate mass. Note that the difference in the fitted masses at $T=47$ and $95$ MeV is $a_\tau m = 0.0009$, corresponding to 5 MeV, indicating both the sensitivity of this analysis and the uncertainty. In the other three channels, the fitted masses at $T=47$ and $95$ MeV are closer, down to 1 MeV in the $D_s^*$ case.

Overall, we conclude that the presence of the plateaus in the ratios indicates that the Ansatz for the spectral function, a narrow peak whose width is negligible, is sufficient, leading to a determination of the groundstate masses at $T=47$ MeV within about 5 MeV, see Table \ref{tab:masses-w-error-T0}. For completeness, we note that our estimates are consistent within errors with those determined by HadSpec at $T= 24$ MeV ($N_\tau=256$) \cite{Cheung:2016bym}. There is also a satisfactory agreement with the PDG values, even though we are not at the physical point and have results at one lattice spacing only.

\begin{figure}[!p]
\begin{center}
\includegraphics[width=0.7\textwidth]{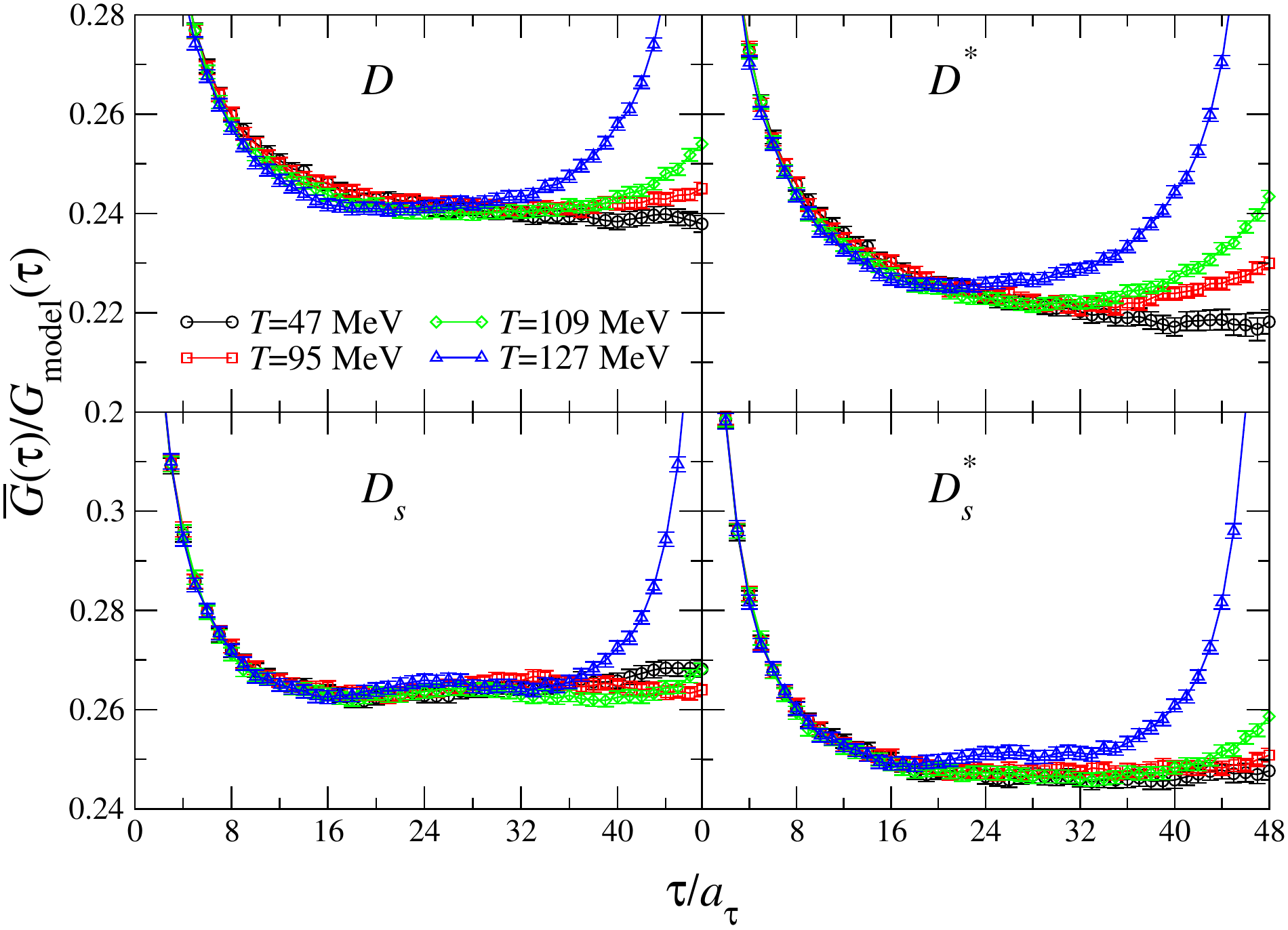} 
\end{center}
  \caption{Ratio $r(\tau;T,T_0)=\overline G(\tau;T)/G_{\rm model}(\tau;T, T_0)$ at the four lowest  temperatures, using as input the groundstate masses determined at $T_0=47$ MeV. 
  }
    \label{fig:G-ratio-model-lowT}
\begin{center}
\includegraphics[width=0.7\textwidth]{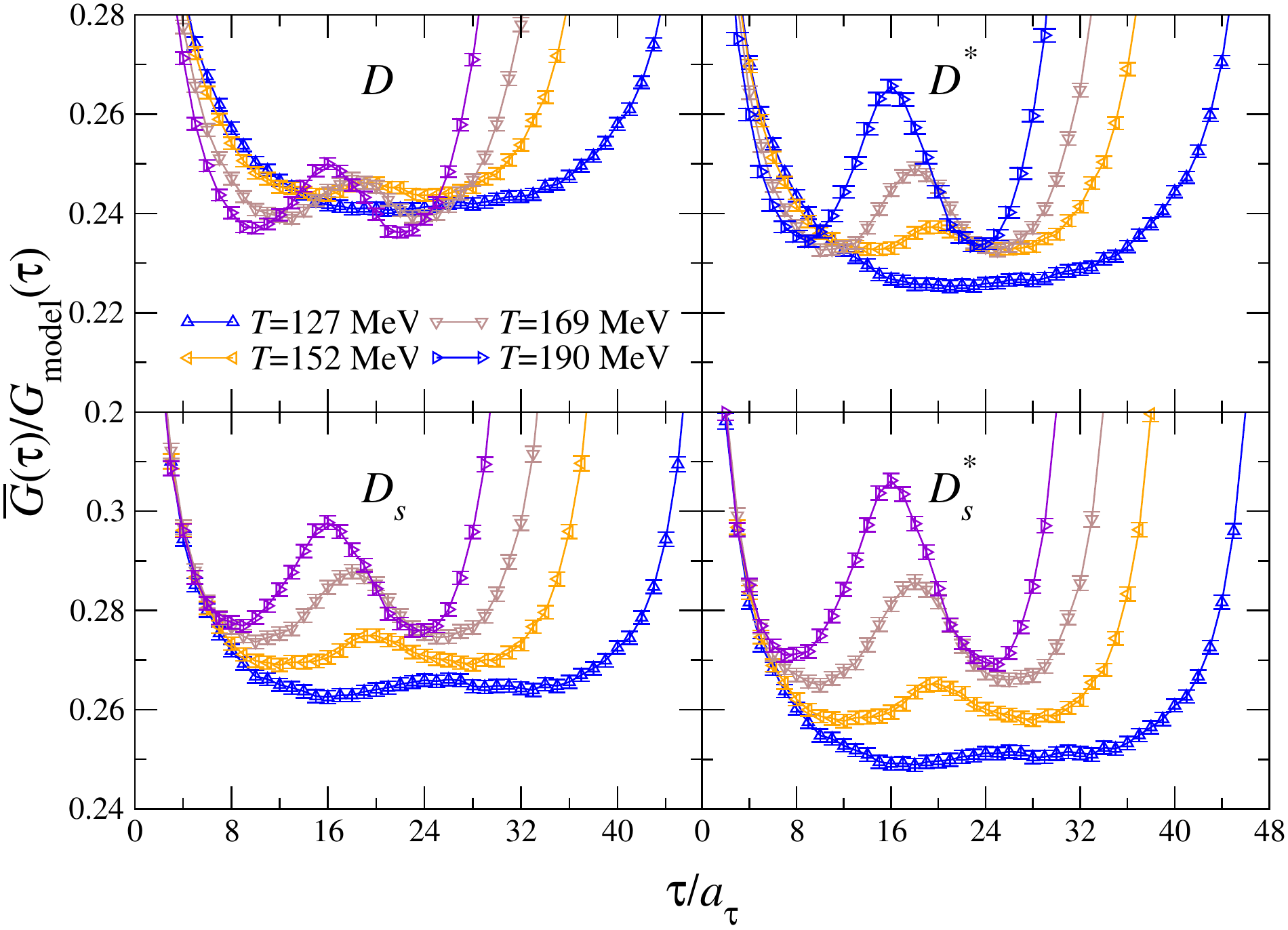} 
\end{center}
  \caption{As above, at four temperatures around the thermal transition.
  }
    \label{fig:G-ratio-model-middleT}
\end{figure}

\subsection{Temperature dependence of correlators}

We now extend this analysis to all temperatures in the hadronic phase as well as just above the transition. We take the ratios of the correlators at temperature $T$ with the model correlator, using the masses determined at $T_0=47$ MeV, see Eq.~(\ref{eq:r}).\footnote{We keep the amplitude $A(T_0)=1$, which does not affect the reasoning.} 
The results are shown in Fig.\ \ref{fig:G-ratio-model-lowT} in the hadronic phase and in Fig.\ \ref{fig:G-ratio-model-middleT} around the crossover. Note that the ratio at $T=127$ MeV is shown in both plots for reference.
 The compactness of the lattice means that at higher temperature the lattice is shorter, but up to $\tau/a_\tau=N_\tau/2$ a comparison between temperatures is possible. 
We note that at the lower temperatures the behaviour is consistent, i.e.\ the correlator ratios are all very similar, except for the reduced extent in the temporal direction. At the highest temperature in the hadronic phase ($T=152$ MeV), a deviation is seen, which gets much stronger in the QGP phase. The wiggles observed around the centre of the lattice at $T=152$ MeV are similar to what is seen in the simple model of Sec.\ \ref{sec:model}, see Fig.~\ref{fig:G-toy} (right), and might indicate a reduction in  mass.  

\begin{figure}[t]
\begin{center}
\includegraphics[width=0.7\textwidth]{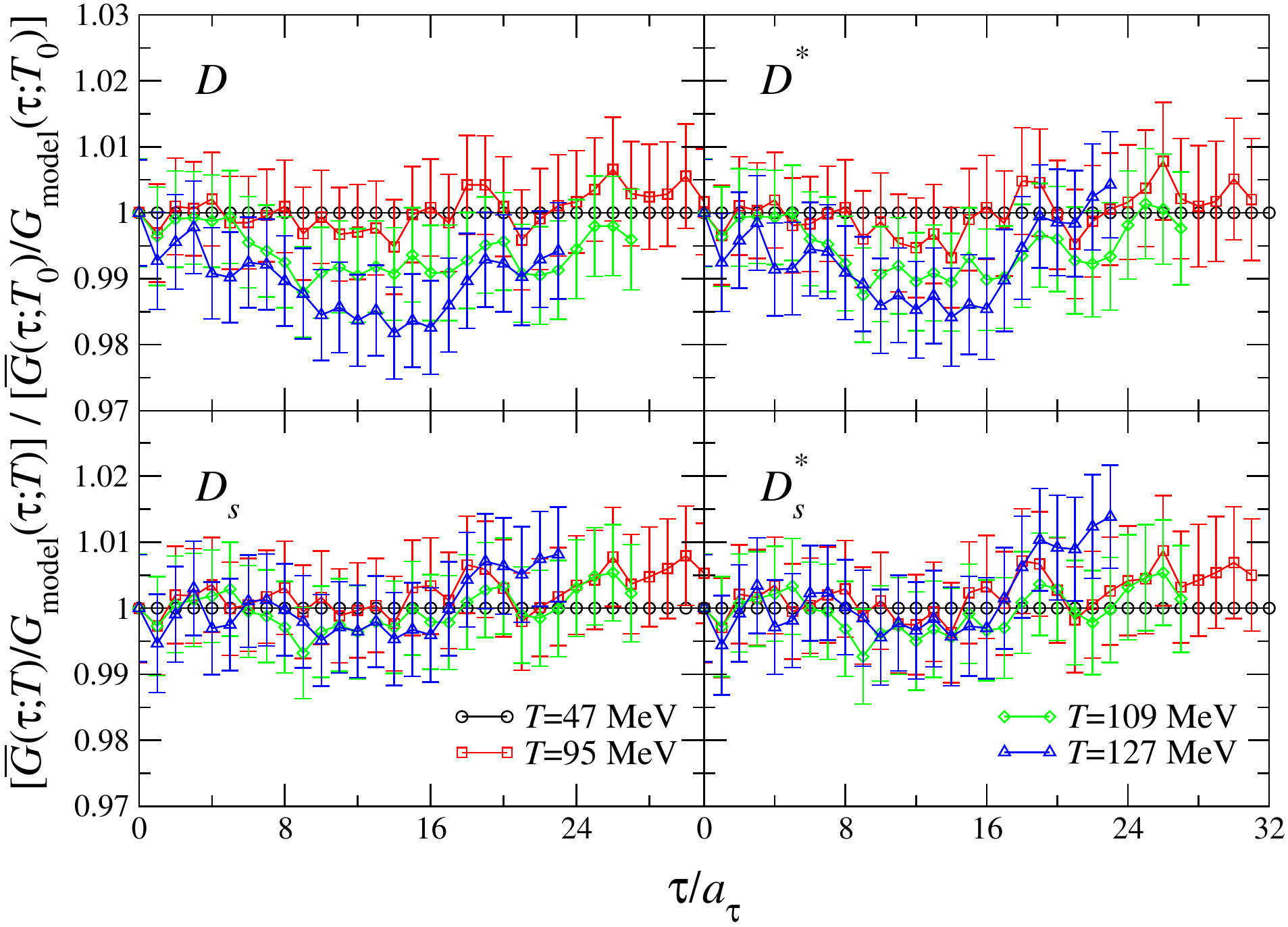} 
\end{center}
  \caption{Double ratio $R(\tau;T,T_0) =${\small $[\overline G(\tau; T)/G_{\rm model}(\tau; T)]/[\overline G(\tau; T_0)/G_{\rm model}(\tau; T_0)]$}, at the four lowest temperatures, using as input the groundstate masses determined at $T_0=47$ MeV. At each temperature, ratios are shown up to $\tau/a_\tau=N_\tau/2$.  
     }
    \label{fig:G-double-ratio-model-lowT}
\end{figure}

\begin{figure}[t]
\begin{center}
\includegraphics[width=0.7\textwidth]{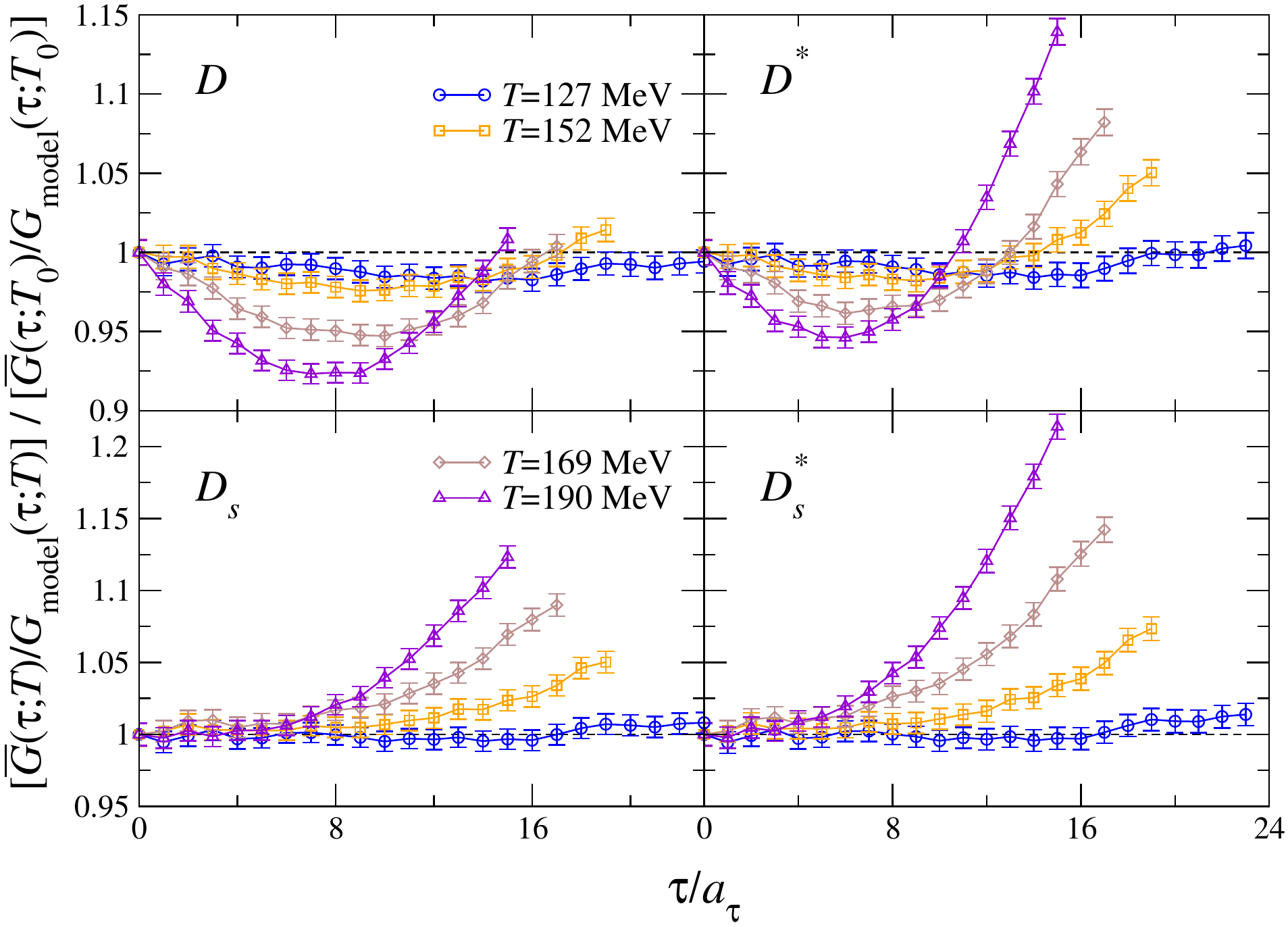} 
\end{center}
  \caption{As above, at four temperatures around the thermal transition.
  }
    \label{fig:G-double-ratio-model-middleT}
\end{figure}

Following Sec.\ \ref{sec:idea} we can take the final step and consider the double ratio in Eq.~(\ref{eq:R}). The results are shown in Fig.~\ref{fig:G-double-ratio-model-lowT} in the hadronic phase and in Fig.~\ref{fig:G-double-ratio-model-middleT} around the crossover (note the disparity in vertical scale between the two plots). Note that the double ratio at $T=127$ MeV is shown in both plots for reference. As mentioned, the only input parameter is $m_0$, the fitted mass at $T_0=47$ MeV. In the double ratio the amplitude $A(T_0)$, see Eq.\ (\ref{eq:Gmodel}), cancels. 
We first observe that this construction indeed eliminates most of the $\tau$ dependence in the hadronic phase. In Fig.\ \ref{fig:G-double-ratio-model-lowT}  the double ratio is essentially consistent with 1, within the error and with deviations of less than 2\%. The most straightforward interpretation is that the spectral content is unchanged at these temperatures.  

Closer to the transition, however, the ratio deviates significantly from 1, as seen in Fig.~\ref{fig:G-double-ratio-model-middleT}. 
At $T=152$ MeV, the deviation is on the order of 5\% (somewhat less in the $D$ channel), rising rapidly in the QGP phase, up to 20\%. The conclusion is 
that at these temperatures the underlying spectral functions have changed in response to the increasing temperature. 
 
\subsection{Groundstate masses in the hadronic phase}

\begin{figure}[t]
\begin{center}
  \includegraphics[width=0.48\textwidth]{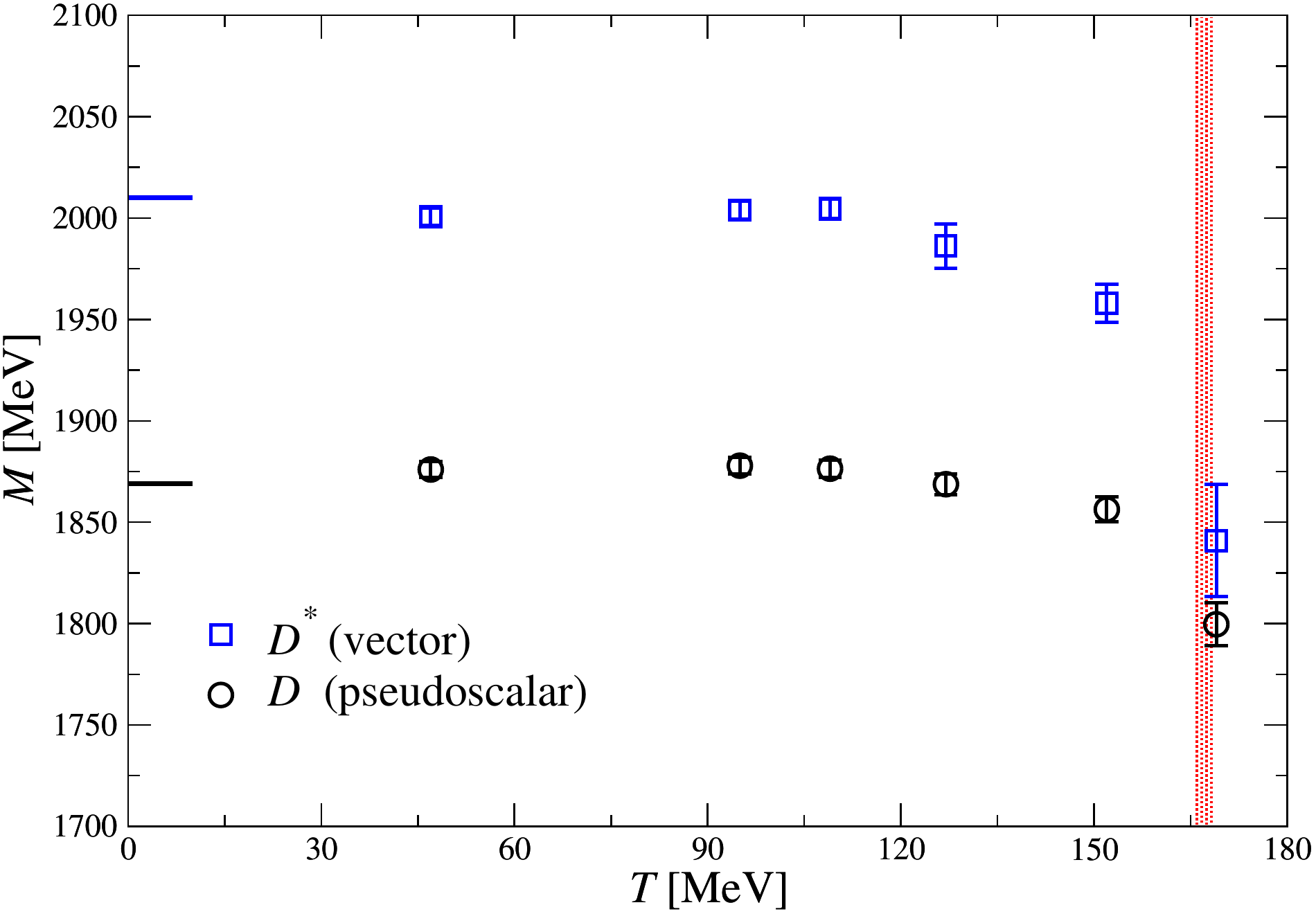} 
  \includegraphics[width=0.48\textwidth]{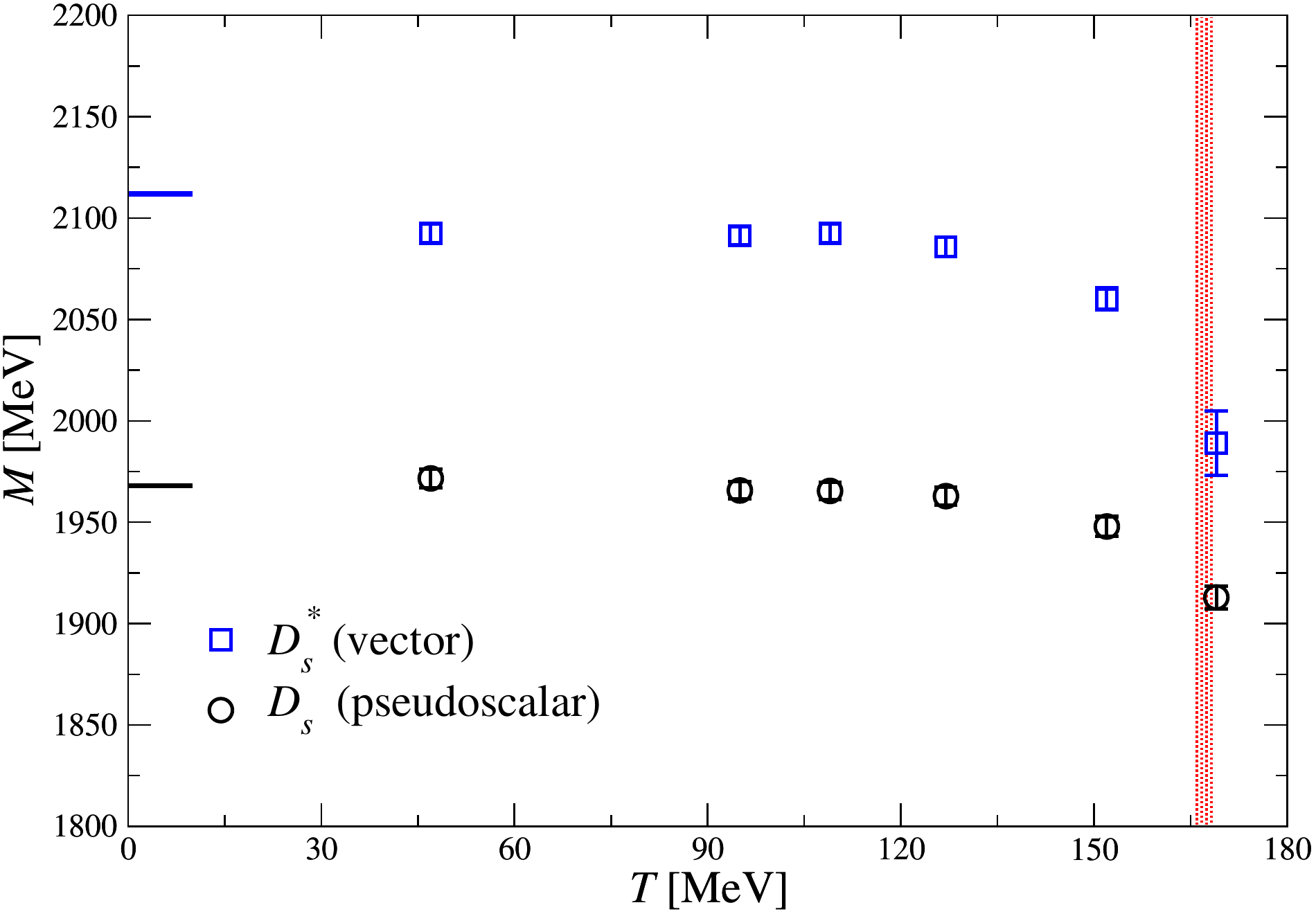}
\end{center}
  \caption{Temperature dependence of the groundstate masses in the hadronic phase, for $D$ and $D^*$ (left) and $D_s$ and $D_s^*$ (right) mesons. The vertical band indicates the thermal transition, while the horizontal stubs at $T = 0$ represent the PDG values 
  \cite{ParticleDataGroup:2020ssz}.
}
    \label{fig:D-T}
\end{figure}

In the preceding subsection, we established that there is little intrinsic temperature dependence in the correlators up to $T=127$ MeV, but that thermal effects are visible at $T=152$ MeV, which quickly get stronger in the QGP. 
To reach this conclusion, only the fitted groundstate mass at $T_0$ was required. Based on this observation, it is justified to also determine the groundstate masses at the other temperatures in the hadronic phase, applying the same fitting procedure as at $T=47$ MeV. The results are given in Table \ref{tab:masses-w-error} 
 and shown in Fig.\ \ref{fig:D-T}. The five lowest temperatures are in the hadronic phase, with the sixth one at (or just above) the crossover region.

 In all four channels, the groundstate mass remains almost constant at low temperature, with a reduction of 20 to 40 MeV only at $T=152$ MeV. At the crossover ($T=169$ MeV), it is still possible to extract a mass, albeit with a larger uncertainty in three of the four channels. We note that the mass splitting between the pseudoscalar and vector channels is about 120 MeV and is only somewhat reduced at the crossover. This is due to a stronger reduction of the vector masses, which is consistent with the stronger temperature dependence observed for the vector channels in the double ratio in Fig.\ \ref{fig:G-double-ratio-model-middleT}.
In the QGP, the change in spectral content found in the preceding subsection no longer justifies the use of a narrow peak to describe the groundstate and hence we refrain from applying the fitting procedure at the higher temperatures.

These results agree with and improve on the previous results of Ref.~\cite{Kelly:2018hsi}, which were obtained with heavier light quarks ($m_\pi=390$\,MeV) and using point sources only.  In that study, significant thermal modifications in the correlators were found for $T\geq141$ MeV, but the spectral function analysis did not reveal any significant mass shift.  Here, using smeared sources we have been able to extract mass shifts in the hadronic phase.

\begin{table}[t]
  \begin{center}
  \resizebox{\textwidth}{!}{
         \begin{tabular}[t]{|c|c|c|cccccc|}
         \hline
	 & $J^P$ & PDG & $T$[MeV]$=47$ & 95 & 109 & 127 & 152 & 169 \\
	\hline
    	 $D$  	& $0^-$ 	& 1869.65(5) 	& $1876(4)$ & $1878(4)$ & $1876(4)$ & $1869(5)$ & $1856(6)$ & $1800(11)$ \\
    	 $D^{*}$  	& $1^-$	& 2010.26(5) 	& $2001(4)$ & $2004(4)$ & $2005(5)$ & $1986(11)$ & $1958(9)$ & $1841(28)$ \\
	 \hline
	 $D_s$  	& $0^-$ 	&  1968.34(7)  	& $1972(5)$ & $1966(4)$ & $1965(4)$ & $1963(4)$ & $1948(5)$ & $1913(6)$ \\
	 $D_s^*$  	& $1^-$  	&  2112.2(4) 	& $2092(4)$ & $2091(5)$ & $2092(5)$ & $2086(5)$ & $2060(6)$ & $1989(16)$ \\
	\hline
   \end{tabular}
   }
    \end{center}
     \caption{$D^{(*)}$ and $D^{(*)}_s$ groundstate masses (in MeV) as a function of temperature in the hadronic phase. The error reflects the combined statistical and systematic uncertainty from the fitting procedure as well as the uncertainty from the scale setting.      }
     \label{tab:masses-w-error} 
\end{table}

We can now compare these results with those found in Refs.\ \cite{Montana:2020lfi,Montana:2020vjg}, obtained using an effective field theory based on chiral and heavy-quark spin-flavour symmetries. In those references it can be seen that between $T=0$ and 150 MeV the $D^{(*)}$ meson masses drop by about 40 MeV and the $D_s^{(*)}$ masses by about 20 MeV. 
In our case,  when comparing the $T=47$ and 152 MeV results we find a reduction in mass of 
 20(7) MeV for the $D$ meson, 
 43(10) MeV for the $D^*$ meson, 
 24(7) MeV for the $D_s$ meson, 
 and 
 32(7) MeV for the $D_s^*$ meson. 
 Note that this is an effect at the percent level and hence satisfyingly similar.
Concerning the width, in Refs.\ \cite{Montana:2020lfi,Montana:2020vjg} all widths are negligible at $T=0$ MeV and increase at $T=150$ MeV to 70 MeV for the $D^{(*)}$ mesons and 20 MeV for the $D_s^{(*)}$ mesons.
Given that $\Gamma/M \ll 1$, this can still be classified as narrow. In our fitting procedure the width is ignored, as already pointed out.

\section{Scalar and axial-vector channels}
\label{sec:SAX}

\subsection{Groundstate masses at the lowest temperatures}

 \begin{figure}[t]
\begin{center}
\includegraphics[width=0.7\textwidth]{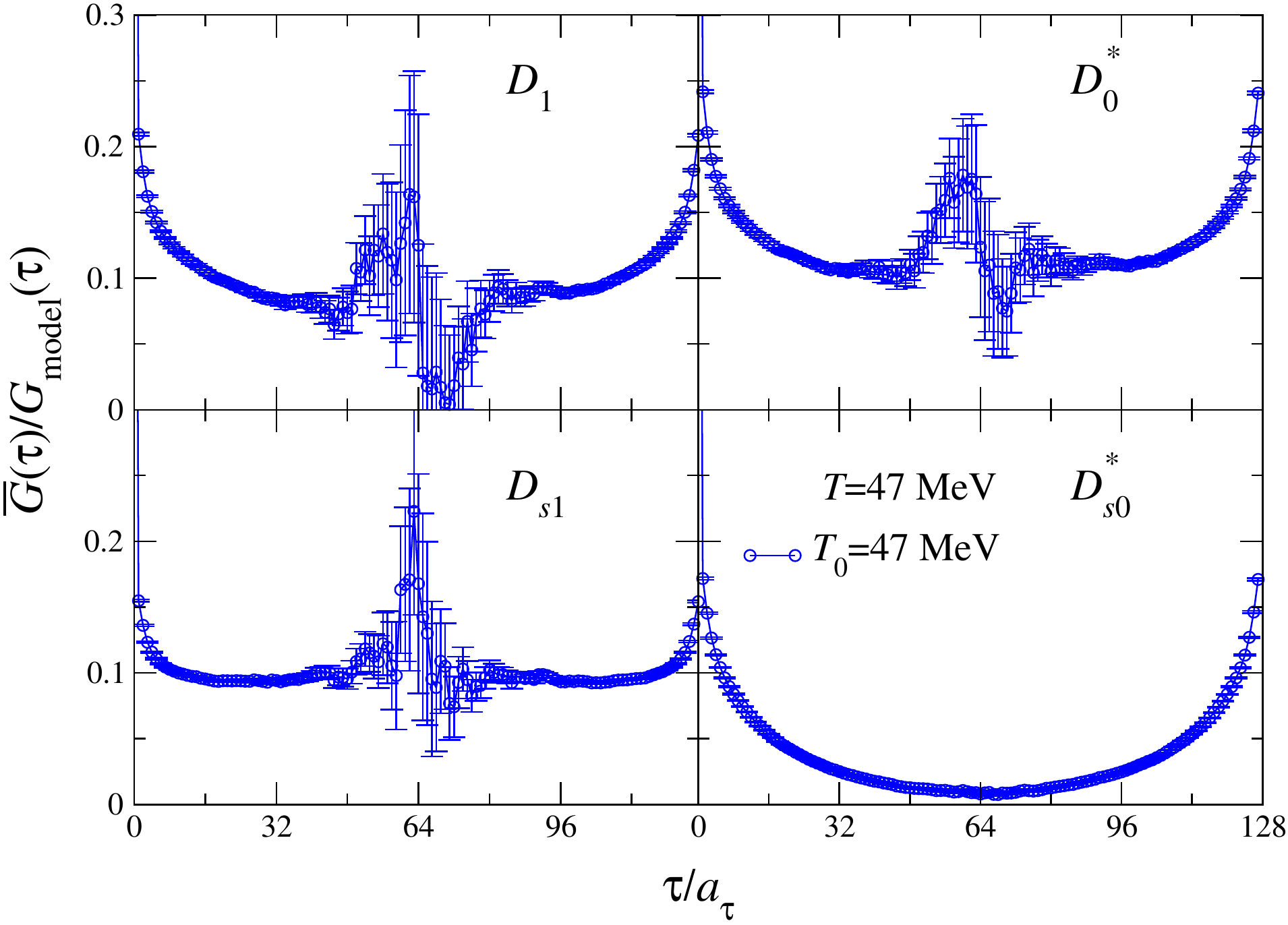} 
\end{center}
  \caption{Ratio $r(\tau;T,T_0)=\overline G(\tau;T)/G_{\rm model}(\tau;T, T_0)$ at $T=47$ MeV,  in the axial-vector and scalar channels, using as input the groundstate masses determined at $T_0=47$ MeV. 
       }
    \label{fig:G-double-ratio-model-AX-SC-128}
\end{figure}

We now turn to the scalar and axial-vector channels. To assess the values found for the groundstate masses at $T=47$ MeV---see Table \ref{tab:masses-w-error-T0}---we consider again the ratio with the model function, see Fig.\ \ref{fig:G-double-ratio-model-AX-SC-128}. 
We observe that the data is much noisier, except in the $D^*_{s0}$ channel. 
Moreover the presence of a plateau in the centre of the lattice is more debatable, except in the $D_{s1}$ channel. This is reflected in the fitting errors included in Table \ref{tab:masses-w-error-T0}.

\begin{figure}[t]
\begin{center}
\includegraphics[width=0.7\textwidth]{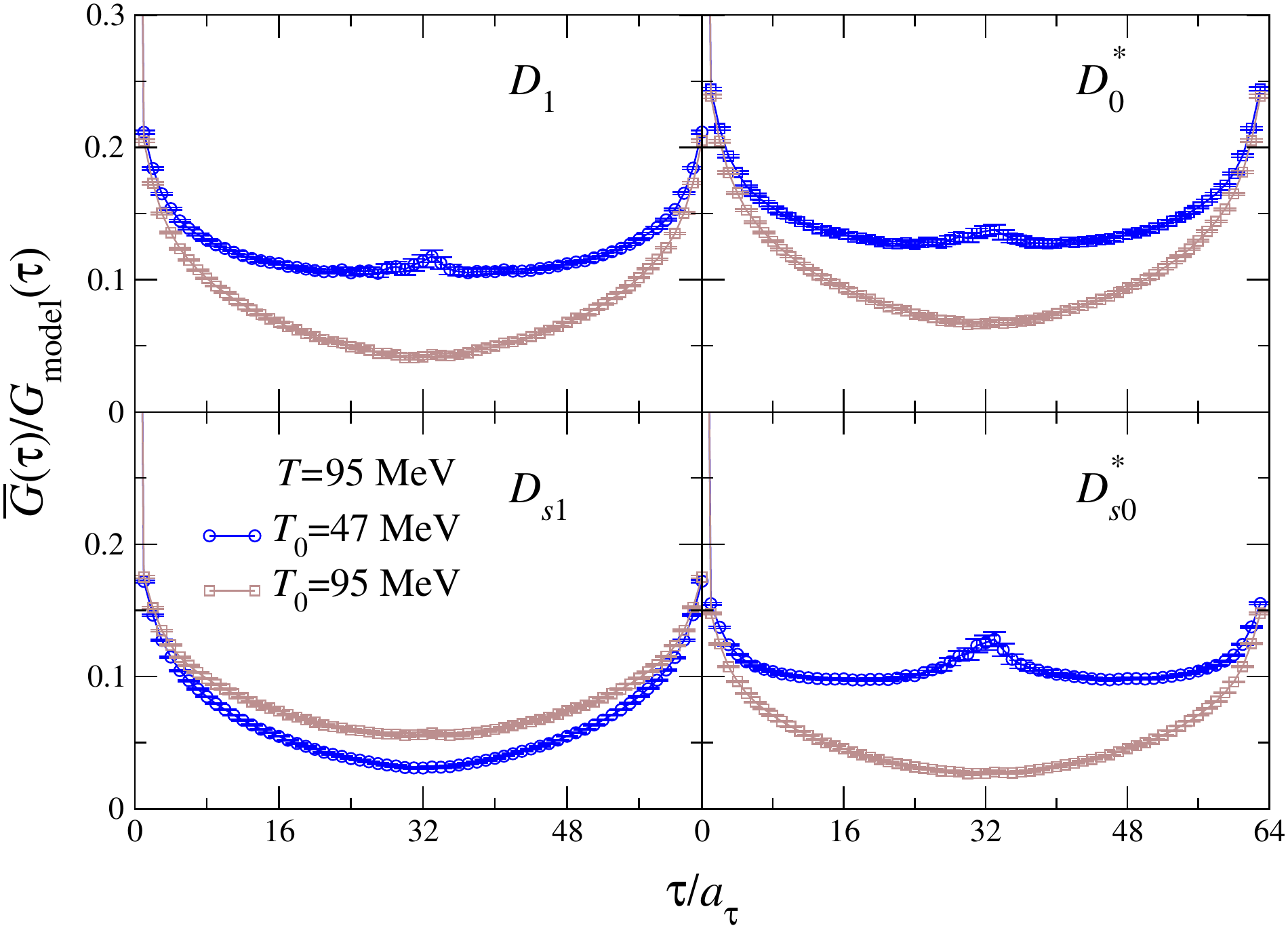} 
\end{center}
  \caption{As above at $T=95$ MeV, using as input the groundstate masses determined at both $T_0=47$ and $T_0=95$ MeV. 
       }
    \label{fig:G-double-ratio-model-AX-SC-64}
\end{figure}

Since for $T=47$~MeV the noise dominates at large Euclidean times, we also consider the correlators at the second-lowest temperature, $T=95$ MeV,
and use for the model mass parameter the fitted masses at both $T_0=47$ and $95$ MeV --  see Table \ref{tab:masses-at}. 
The ratios are shown in Fig.\ \ref{fig:G-double-ratio-model-AX-SC-64}. While the noise is indeed drastically reduced, we also note the absence of a plateau, in all four channels. The cusp-like behaviour for the ratio with $T=47$ MeV masses is an indication that the model mass is too large -- see Sec.\ \ref{sec:model}. 

Overall, however, the conclusion is that the simple fitting Ansatz cannot be justified, as indicated by the absence of a plateau; we discuss possible reasons for this below. In comparison to the  vector and pseudoscalar channels, the fitted masses for (most of) the scalar and axial vector channels have somewhat larger errors, and compare less favourably with the PDG values (Tab.~\ref{tab:masses-w-error-T0}). The $D_{s0}^*$ state, in particular, is anomalously light, and notably has a mass which is consistent with our $D_s^*$.

\begin{figure}[t]
\begin{center}
\includegraphics[width=0.7\textwidth]{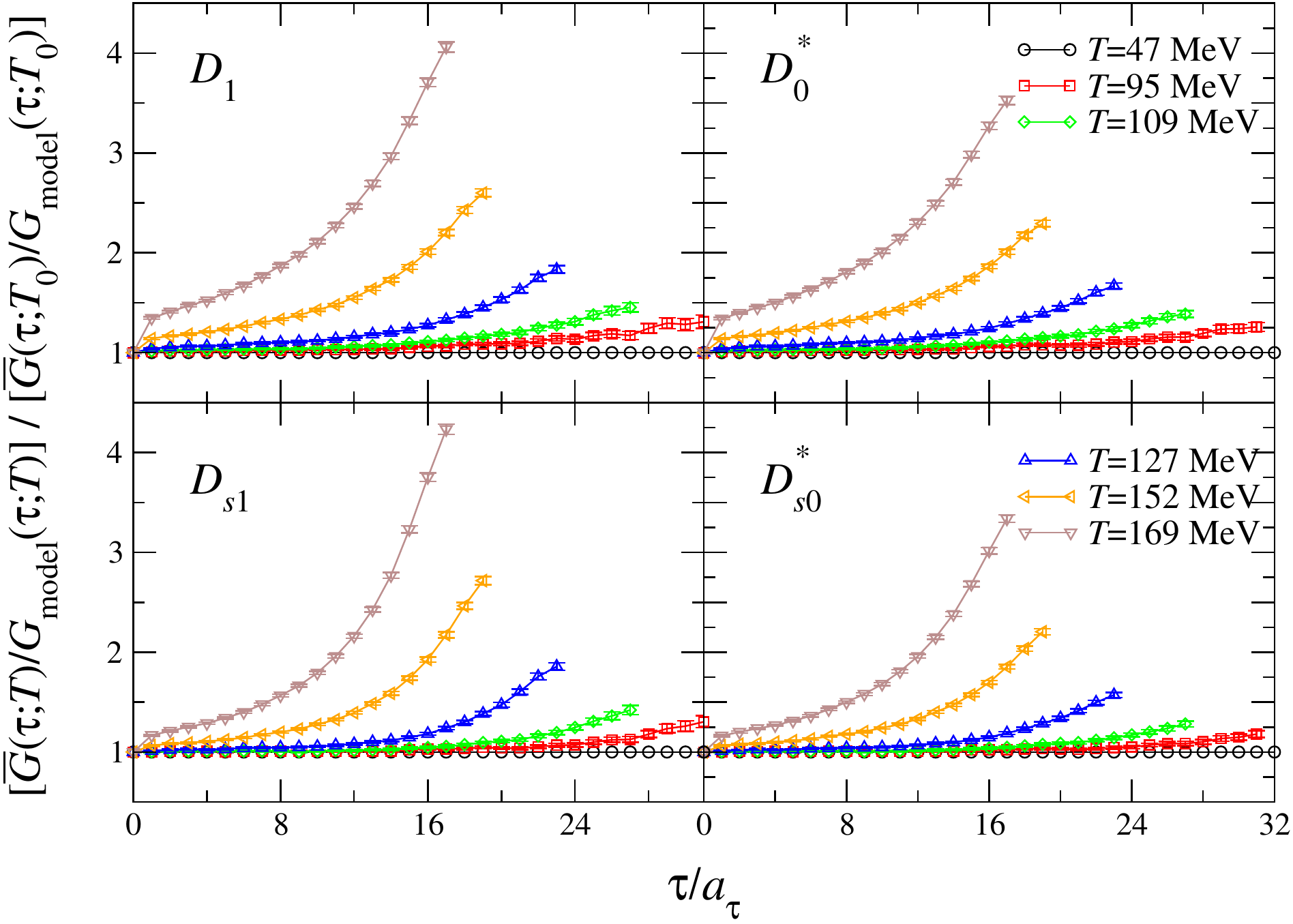} 
\end{center}
  \caption{Double ratio $R(\tau;T,T_0) =${\small $[\overline G(\tau; T)/G_{\rm model}(\tau; T)]/[\overline G(\tau; T_0)/G_{\rm model}(\tau; T_0)]$}, 
  in the axial-vector and scalar channels, using as input the groundstate masses determined at $T_0=47$ MeV only. At each temperature, ratios are shown up to $\tau/a_\tau=N_\tau/2$.  
     }
    \label{fig:G-double-ratio-model-AX-SC}
\end{figure}

\subsection{Temperature dependence of correlators} 

To analyse the temperature dependence, we consider again the double ratio (\ref{eq:R}), with the results shown in Fig.\ \ref{fig:G-double-ratio-model-AX-SC}. We observe a much stronger temperature dependence throughout the hadronic phase than before. To clarify that this is not entirely due to the mismatch of the model and actual correlator, it is useful to consider the double ratio as in Eq.~(\ref{eq:R2}), i.e.\ as the ratio of lattice correlators over the ratio of model correlators.  It is easy to see that the latter is much closer to 1 than the double ratios shown in Fig.\ \ref{fig:G-double-ratio-model-AX-SC}, and hence the strong temperature effect is indeed due to the lattice correlators themselves.
This is further supported by the analysis of reconstructed correlators, see App.\ \ref{sec:G-rec}.\footnote{As a side note, we remark here that also in quarkonia P wave states show a much stronger temperature dependence than S wave states, see e.g.\ Ref.\ \cite{Aarts:2014cda}.} Given these results we refrain from using the fitting procedure  in these channels, as it is not justified.

\subsection{Discussion}

The failure of the fitting Ansatz in the scalar and axial vector channels is not entirely unexpected. Indeed, the main issue hindering a further understanding of thermal effects is that the states considered here are not well described by narrow peaks, nor by the local operators employed in the simulation, already at vanishing temperature.
The difficulty is that the quantum numbers of these states imply that they couple to nearby two-meson thresholds in S-wave. The impact of such thresholds on hadron properties is clear in the zero-temperature data: $D_{s0}^*$ and $D_{s1}$ are below $D K$ and $D^*K$ threshold, and as such have narrow widths, whereas $D_0^*$ and $D_1$ are above $D\pi$ and $D^*\pi$ threshold, and so are very broad resonances. Note that our lattice results for the pion and kaon groundstate masses are given in Table \ref{tab:masses-w-error-pi-K}.\footnote{The pion is heavier than in nature, due to the unphysical light quarks in the lattice simulation, but the kaon is close to its physical value.} 
On this basis, we may already anticipate the failure of our simple Ansatz (which assumes zero width) for the $D_0^*$ and $D_1$ states, but also possibly for $D_{s0}^*$ and $D_{s1}$, since it is not automatic that such states would remain bound with respect to their corresponding thresholds with our lattice parameters. Indeed, as a curiosity, we notice that the $D_{s1}$ state extracted from our fitting procedure (at $T=47$~MeV) has a mass which is consistent with the $D^*K$ threshold.

As well as imparting a width on states above threshold, the coupling to thresholds confounds the extraction of hadron masses in a more fundamental way, as exemplified by lattice studies of hadron resonances in the L\"uscher method. This is discussed in detail in Ref.~\cite{Gayer:2021xzv} where the energy dependence of elastic $D\pi$ scattering is mapped out and a $D_0^\ast$ resonance pole is identified (including at the pion mass used in this work). In a similar analysis~\cite{Cheung:2020mql} the $D_{s0}^\ast$ state was computed and found to be significantly heavier than the $D_0^\ast$ state, in contrast to early experimental studies of these systems. The pole couplings of the $D_0^\ast$ and $D_{s0}^\ast$ poles to $D\pi$ and $DK$ respectively were found to be compatible suggesting a similar origin, while the difference in widths between the $D_{s0}^\ast$ and the $D_0^\ast$ may result from the very different phase space available in each case.

\begin{figure}[t]
\begin{center}
\includegraphics[width=0.49\textwidth]{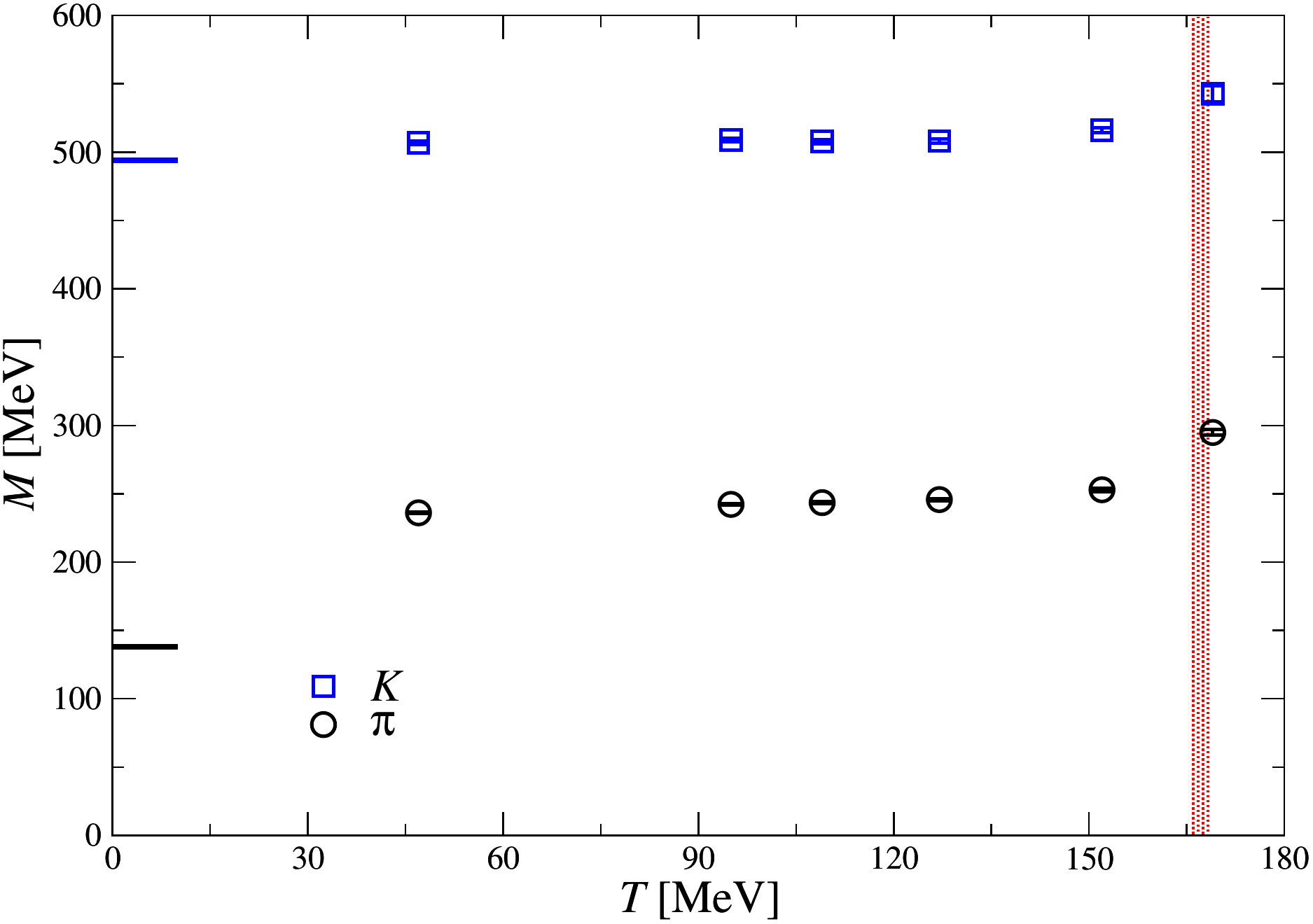} 
\end{center}
\caption{As in Fig.\ \ref{fig:D-T}, for the 
pion and kaon groundstate masses in the hadronic phase. Note that the pion is heavier than in nature, due to the unphysical light quarks in the lattice simulation. 
       }
    \label{fig:M-pi-K}
\end{figure}

Extending such analyses to finite temperature would be very interesting and a first step to do so would be to implement the larger set of operators used for studies in vacuum also here. In addition to expected changes in hadron masses and widths as a function of temperature, more drastic possibilities also arise, including transitions from bound states to resonances, or vice versa.
For this reason we show in Table \ref{tab:masses-w-error-pi-K} and Fig.~\ref{fig:M-pi-K} the temperature dependence of the pion and kaon groundstate masses. We observe that both $\pi$ and $K$ masses increase slightly as the crossover is approached, whereas the
$D^{(*)}$ and $D_s^{(*)}$ masses decrease. 
This may result in an interesting interplay between these states as the temperature increases, which is an interesting direction for further study.

\begin{table}[t]
  \begin{center}
  \resizebox{\textwidth}{!}{
         \begin{tabular}[t]{|c|c|c|cccccc|}
         \hline
	 & $J^P$ & PDG & $T$[MeV]$=47$ & 95 & 109 & 127 & 152 & 169 \\
	\hline
    	 $\pi$  	& $0^-$ 	& 139.57039(18)   & 239(1) & 242(1) & 243(1) &  246(1) &  253(1) &  295(2) \\
    	 $K$  	& $0^-$	& 493.677(16) & 507(1) & 509(1) & 508(1) & 508(2) & 516(2) & 543(6) \\
	\hline
   \end{tabular}
   }
    \end{center}
     \caption{As in Table \ref{tab:masses-w-error}, for the pion and kaon  groundstate.
         }
     \label{tab:masses-w-error-pi-K} 
\end{table}

\section{Conclusion}
\label{sec:conclusion}

We have presented a systematic study of $D$ and $D_s$ mesons throughout the hadronic phase, using lattice QCD. By constructing ratios of correlators, we conclude that there is essentially no temperature dependence in the $D_{(s)}$ and $D_{(s)}^*$  (pseudoscalar and vector) channels, at least up to $T=127$ MeV in our simulations. This absence of thermal effects is quite robust, as it does not rely on a particular Ansatz for spectral content, except at the lowest temperature where a simple narrow spectral function for the groundstate is assumed and also justified. This method can be used more widely to asses lattice QCD data. We benefit from using our FASTSUM anisotropic lattices which are particularly suited for spectroscopy.
Thermal effects become noticeable at $T=152$ MeV and are described by a reduction of the groundstate masses, by 20 to 40 MeV. This is the paper's main result in the context of thermal QCD phenomenology and can e.g.\ be used to benchmark effective model descriptions used to describe heavy-ion data.

In the case of the $D_{1(s)}$ and $D_{0(s)}^*$  (axial-vector and scalar) channels, a very different picture emerges.
The description of the groundstate in terms of a narrow spectral function is less justified and a strong temperature dependence throughout hadronic phase is observed. Here a better understanding of correlators and spectral functions is required, e.g.\ as provided by hadronic models. On the lattice, it might be of interest to carry out an analysis with an extended operator set, as is done in vacuum, also at finite temperature. As in-medium masses change with temperature close to the crossover, this may lead to subtle threshold effects, which would be interesting to observe.

\vspace{0.5cm}
\noindent
{\bf Acknowledgements} --
We thank Christopher Thomas and David Wilson for discussion.
 This work is supported by the UKRI Science and Technology Facilities Council (STFC) Consolidated Grant No.\ ST/T000813/1 and 
 by STRONG-2020 ``The strong interaction at the frontier of knowledge: fundamental research and applications'' which received funding from the European Union's Horizon 2020 research and innovation programme under grant agreement No 824093.
 S.~K.\ is supported by the National Research Foundation of Korea under Grant No.\ NRF-2021R1A2C1092701 funded by the Korean government (MEST) and in part by NRF-2008-000458. 
 We are grateful to DiRAC, PRACE and Supercomputing Wales for the use of their computing resources and to the Swansea Academy for Advanced Computing for support. 
 This work was performed using the PRACE Marconi-KNL resources hosted by CINECA, Italy and the DiRAC Extreme Scaling service and Blue Gene Q Shared Petaflop system at the University of Edinburgh operated by the Edinburgh Parallel Computing Centre. The DiRAC equipment is part of the UK’s National e-Infrastructure and was funded by UK’s BIS National e-infrastructure capital grant ST/K000411/1, STFC capital grants ST/H008845/1 and ST/R00238X/1, and STFC DiRAC Operations grants ST/K005804/1, ST/K005790/1 and ST/R001006/1.

\el

\noindent
{\bf Open Access Statement} -- For the purpose of open access, the authors have applied a Creative Commons Attribution (CC BY) licence to any Author Accepted Manuscript version arising.

\appendix

\section{Lattice details}
\label{sec:lattice}

In this appendix, we give some more details of the lattice simulations, which are fully described in Ref.\ \cite{Aarts:2020vyb} and based on the anisotropic lattice work of the HadSpec collaboration at $T=0$ \cite{Edwards:2008ja,HadronSpectrum:2008xlg}. 
The openQCD-FASTSUM package \cite{glesaaen_jonas_rylund_2018_2216355,glesaaen_jonas_2018_2217028} was used for ensemble generation and computation of correlation functions.
The lattice action consists of a Symanzik-improved gauge action and a Wilson tadpole-improved clover fermion action, with stout-smeared links. The strange quark is at its physical value, but the light quarks are heavier than in nature. This mostly affects the pion mass. 
 Relevant quantities and temperatures available are given in Tables  \ref{tab:lattice_spacings} and \ref{tab:Gens}.

\begin{table}[h]
  \begin{center}
    \begin{tabular}[t]{ | c | c | c | c | c | c | }
	\hline
	$a_\tau$ [fm] 	& $a_\tau^{-1}$ [GeV]  	& $\xi=a_s/a_\tau$ 	& $a_s$ [fm] 	& $m_\pi$ [MeV] & $T_{\rm pc}^{\bar\psi\psi}$ [MeV] \\
	\hline
	 0.03246(7) 	& 6.079(13) 		& 3.453(6) 		& 0.1121(3) 	& 239(1) 	& 167(2)(1) \\
	\hline
  \end{tabular}
  \end{center}
\caption{Details of the Generation 2L ensembles on $32^3\times N_\tau$ lattices. The temporal lattice spacing is determined using the mass of $\Omega$ baryon. $\xi$ is the renormalised anisotropy, determined via the slope of the pion dispersion relation. The pseudocritical temperature $T_{\rm pc}^{\bar\psi\psi}$ is determined via the inflection point of the renormalised chiral condensate \cite{Aarts:2020vyb}.  
The small difference with the numbers listed in Ref.\ \cite{Aarts:2020vyb} is discussed in the text.}
  \label{tab:lattice_spacings}
  \begin{center}
         \begin{tabular}[t]{|c|c|c|c|c|c|c|c|c|c|c|}
	\hline
    	$N_\tau$  		& 128   & 64 	& 56     & 48     & 40     & 36    & 32     & 28     & 24     & 20 \\
	$T$ [MeV]  	&   47   & 95 	& 109   & 127   & 152   & 169  & 190   & 217   & 253   & 304 \\
	$N_{\rm cfg}$  	& 1024 & 1041 	& 1042 & 1123 & 1102 & 1119 & 1090 & 1031 & 1016 & 1030 \\
	\hline
   \end{tabular}
    \end{center}
     \caption{Temporal extent, temperature in MeV, and number of configurations available in Generation 2L. } 
     \label{tab:Gens} 
\end{table}

 The most important update compared to  Ref.\ \cite{Aarts:2020vyb} is an improved estimate of the lattice spacing, as determined by HadSpec \cite{Wilson:2019wfr}.\footnote{We are grateful to Christopher Thomas and David Wilson for discussions on this point.} In Ref.\ \cite{Wilson:2019wfr} the inverse temporal lattice spacing is given as $1/a_\tau = 6079(13)$ MeV, using the computed value of the $\Omega$ baryon mass, $a_\tau m_\Omega=0.2751(6)$.\footnote{Since in nature  $m_\Omega= 1672.45(29)$ \cite{ParticleDataGroup:2020ssz}, the error in the lattice spacing is dominated by the uncertainty of the $\Omega$ mass as computed on the lattice, not by the experimental uncertainty.} The inverse lattice spacing is hence slightly larger than the one determined in earlier HadSpec work \cite{Wilson:2015dqa} and employed in Ref.\ \cite{Aarts:2020vyb},  namely $1/a_\tau = 5997(34)$ MeV. This difference is due to using 64 rather than 16 distillation vectors. As a result, the values (in MeV) in our paper \cite{Aarts:2020vyb}, introducing the so-called Generation 2L ensembles, are all slightly lower than the ones given here, by a factor of $6079/5997 = 1.014$ (or 1.4\%). For the light hadron spectrum this is an effect of a few MeV at most; for $D$ mesons, with masses on the order of 2 GeV, the effect of changing the lattice spacing is around 30 MeV and hence notable.

The improved lattice spacing also results in a slightly higher value of the pion mass, compared to the one quoted in Ref.\ \cite{Aarts:2020vyb}. HadSpec uses the value of  $m_\pi=239$ MeV, based on a finite-volume spectral analysis \cite{Wilson:2014cna}, using a large set of operators, on a lattice with extent $N_\tau=256$.   
Our own determination, using smeared sources of the local operator $O\sim \bar\psi \gamma_5\psi$ only and the procedure outlined in \ \ref{sec:fit}, is $m_\pi=236.0(5)$, obtained on the $N_\tau=128$ lattice. This small discrepancy is not further relevant for the understanding of the thermal effects discussed in this paper. We will follow HadSpec and use $m_\pi=239$ MeV to identify the Generation 2L ensembles.

The charm quark uses the same relativistic action as for the light and strange quarks with the charm-quark mass tuned to reproduce the physical $\eta_c$ mass and the anisotropy parameters tuned to produce a relativistic dispersion relation with appropriate anisotropy. This process is described in Refs.\ \cite{HadronSpectrum:2012gic,Cheung:2016bym}.

\begin{table}[t]
  \begin{center}
  \resizebox{\textwidth}{!}{
         \begin{tabular}[t]{|c|c|cccccc|}
         \hline
	 & $J^P$ & $N_\tau=128$ & 64 & 56 & 48 & 40 & 36 \\
	\hline
   	 $D$  	& $0^-$ & $0.3086(1)$ & $0.3089(1)$ & $0.3087(2)$ & $0.3074(6)$ & $0.3054(8)$ & $0.2960(16)$ \\	 
    	 $D^{*}$  	& $1^-$ & $0.3291(1)$ & $0.3297(1)$ & $0.3298(5)$ & $0.3267(17)$ & $0.3221(14)$ & $0.3028(45)$ \\	 
 	 $D_0^*$	& $0^+$ & $0.3656(14)$ &  $0.3432(24)$ &&&&\\ 
    	 $D_1$  	& $1^+$ & $0.3823(70)$  & $0.3508(59)$ &&&& \\
	 \hline
	 $D_s$  	& $0^-$ & $0.3243(3)$ & $0.3234(1)$ & $0.3233(1)$ & $0.3229(2)$ & $0.3204(4)$ & $0.3147(7)$ \\	 
	 $D_s^*$  	& $1^-$ & $0.3442(1)$ & $0.3440(1)$ & $0.3442(2)$ & $0.3431(3)$ & $0.3389(6)$ & $0.3272(25)$ \\
 	 $D_{s0}^*$ & $0^+$ & $0.3479(46)$ & $0.3666(56)$ &&&&\\ 
	 $D_{s1}$ 	& $1^+$ & $0.4132(2)$	&  $0.365(100)$ &&&&\\ 
	\hline
   \end{tabular}
   }
    \end{center}
     \caption{$D$ and $D_s$ meson masses in temporal lattice units, $a_\tau M$, as a function of the temporal extent of the lattice, $N_\tau$, in the hadronic phase. Errors combine the statistical and systematic uncertainty from the fitting procedure but do not include other possible systematic effects. In case of the scalar and axial-vector channels, results at smaller $N_\tau$ (higher temperature) are not shown, as discussed in the main text.   }     
     \label{tab:masses-at} 
\end{table}

Concerning the quark propagators, we consider local operators of the type $\bar\psi\Gamma c$, where $\psi=\{u, s\}$. In the computation we use both local and smeared sources. For the latter, Gaussian smearing is applied at the source $\left(\eta\right)$ and sink using
\be
\eta^\prime = C\left(1 + \kappa H\right)^{n}\eta,
\ee
where $\eta$ is the bare (local) source, $H$ the spatial hopping part of the Dirac operator, $C$ an appropriate normalisation, and $\kappa$ and $n$ determine the amount of smearing. Here $n=100$ and $\kappa = 5.5$ are used; these parameters were chosen quite some time ago \cite{Aarts:2015mma} such that the zero-temperature nucleon correlator displays good groundstate isolation.
We have analysed correlators put together using local and smeared sources. The results obtained for the groundstates  are consistent, but with a tendency for smeared sources to give a slightly lower value for the groundstate mass.
Overall, we found that masses from local and smeared sources are determined up to a few MeV, with a maximal uncertainty of up to 10 MeV only in some channels.
The extracted $D_{(s)}$ masses (using smeared sources) are given in Table \ref{tab:masses-at} in lattice units. These have been obtained following the procedure explained in App.\ \ref{sec:fit}.

\section{Reconstructed correlators}
\label{sec:G-rec}

Reconstructed correlators can be obtained, without explicit spectral reconstruction, using the identity for hyperbolic functions \cite{Ding:2012sp}
\be
\frac{\cosh\left[\Omega(t-N/2)\right]}{\sinh(\Omega N/2)}
 = \sum_{n=0}^{m-1}\frac{\cosh\left[\Omega(t+ nN-mN/2)\right]}{\sinh(\Omega m N/2)},
\ee
where $t,n,m,N$ are all positive integers.
Introducing the lattice spacing $a_\tau$ and identifying the temperatures $T$ and $T_0$ with the number of points in the time direction, 
\be
T = 1/(a_\tau N), \qquad\qquad T_0 = 1/(a_\tau mN),
\ee
such that
\be
\frac{T}{T_0} = m \in \mathbb{N},
\ee
this identity relates the kernels at temperatures $T$ and $T_0$,  
\be
K(\tau, \om; T) = \sum_{n=0}^{m-1} K(\tau+ a_\tau nN, \om; T_0).
\ee
Inserting this in the spectral relation for a correlator at temperature $T_0$ yields an expression for a correlator at a higher temperature $T=mT_0$, assuming the spectral content is unchanged, i.e.\ the reconstructed correlator
\be
G_{\rm rec}(\tau; T,T_0) = \sum_{n=0}^{m-1} G(\tau+n/T;T_0).
\label{eq:Grec-direct}
\ee
Note that in principle $T/T_0$ should be an integer; in practice this can be avoided by adding data points to the correlator at the lowest temperature, until an integer ratio is obtained. The data points (which for heavy mesons we can take to have the value 0) are added in the centre of the lattice, where the correlator is exponentially suppressed, and hence this has a negligible effect.

\begin{figure}[t]
\begin{center}
\includegraphics[width=0.465\textwidth]{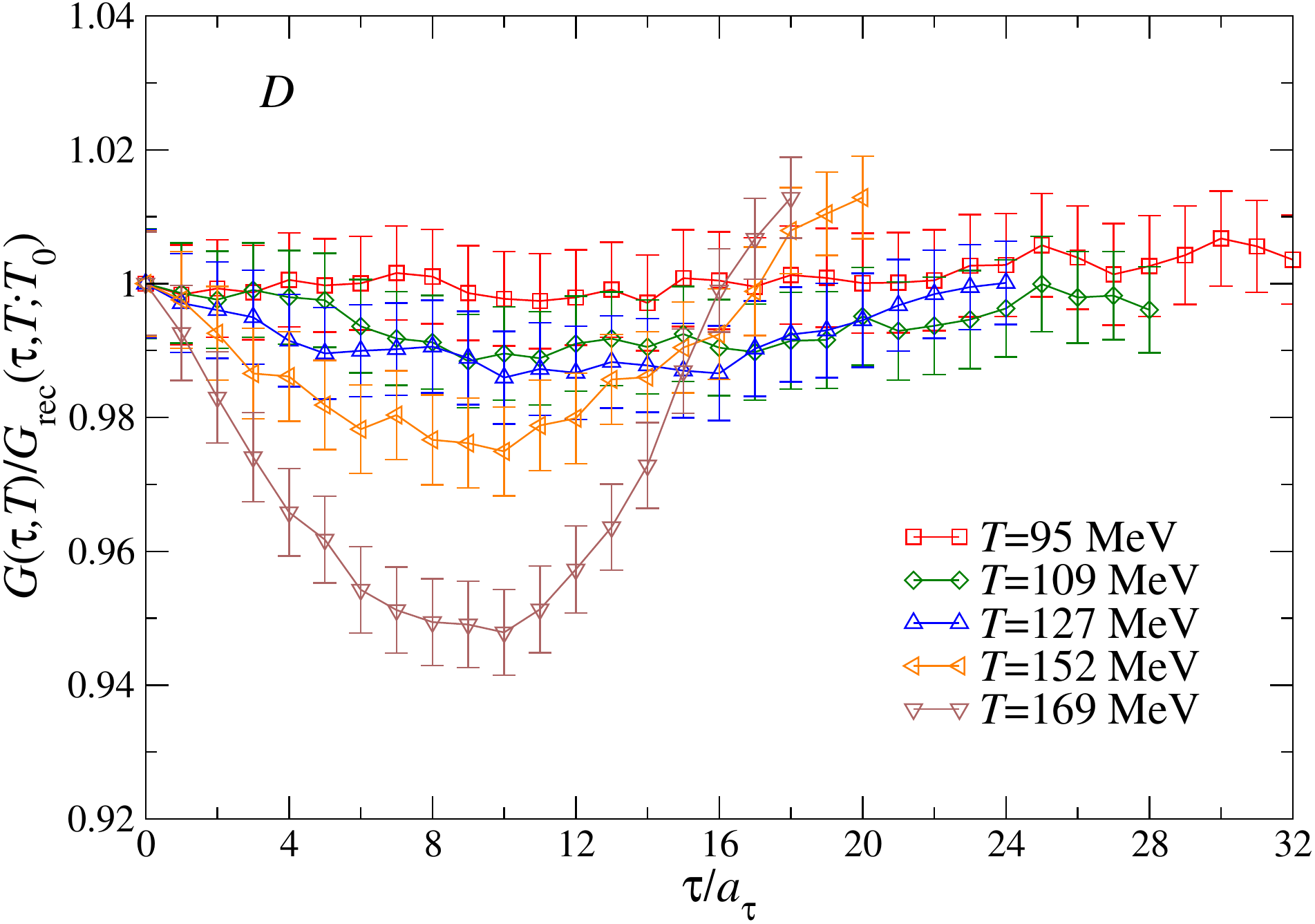}
\includegraphics[width=0.45\textwidth]{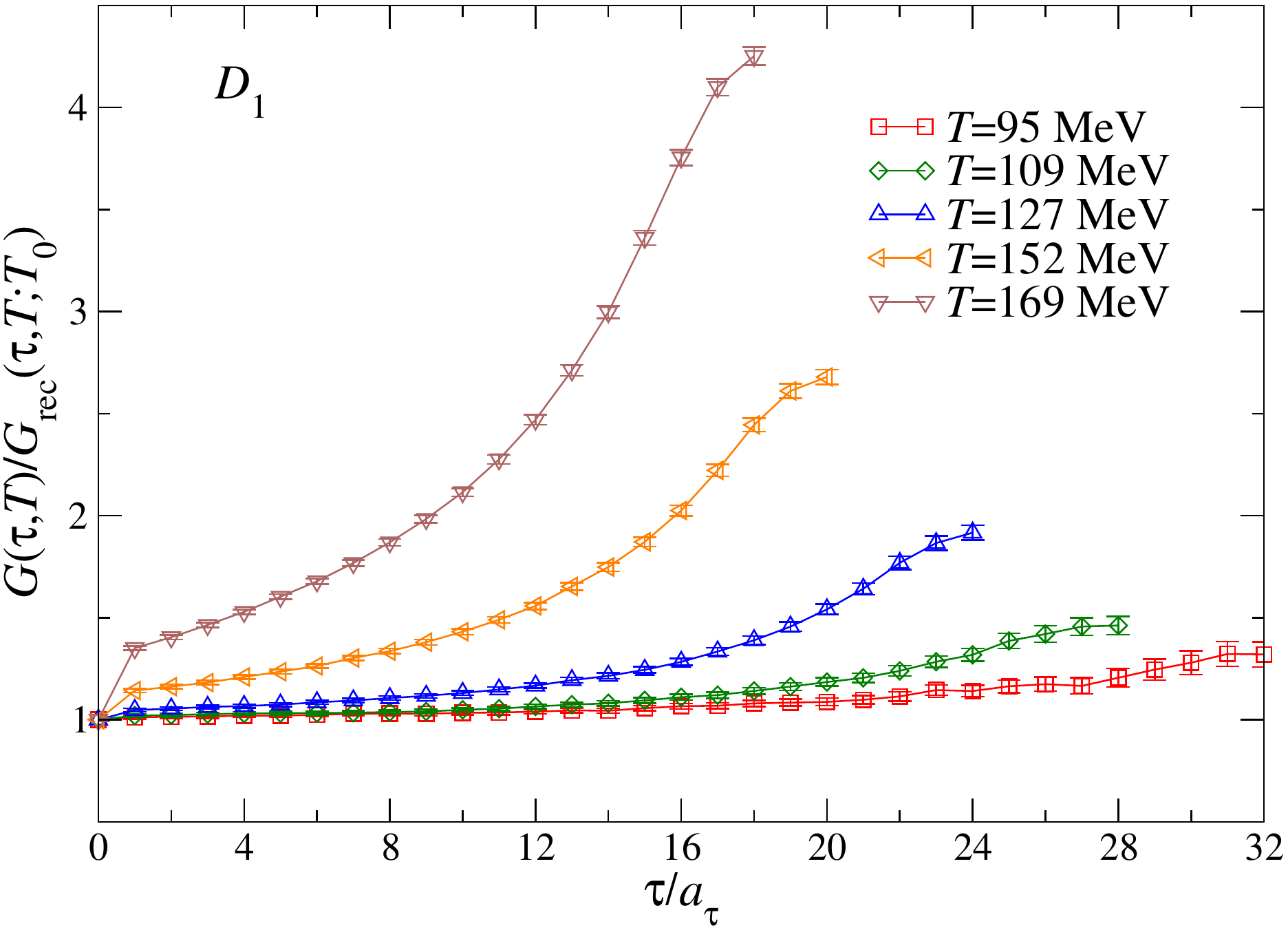}
\end{center}
  \caption{Ratio of correlators and reconstructed correlators in the hadronic phase, using as reference temperature $T_0=47$ MeV, in the $D$ (left) and $D_1$ (right) channel.
  }
    \label{fig:recon}
\end{figure}

Two examples of ratios of correlators at temperature $T$ and reconstructed correlators, with $T_0=47$ MeV, are shown in Fig.\ \ref{fig:recon}. We find the same behaviour as in the main part of the paper for the double ratios: in pseudoscalar and vector channels, temperature effects are not visible within errors up to at least $T=127$ MeV, while in scalar and axial-vector channels, temperature effects are strong throughout the hadronic phase. The size and shape of the ratios is very similar to the ones seen in the double ratios, adding further confidence to the analysis.

\section{Groundstate mass extraction} 
\label{sec:fit}

\subsection{Formulation of the problem}

In this section, we describe the regression-based analysis employed to extract an estimate
of the groundstate mass contributing to a correlation function. Our methodology is
based on the one presented in Ref.\ \cite{Bazavov:2019www}. It can be classified as a nonlinear multi-model
regression analysis in which the contribution of each independent state to the correlation function
is modelled using\footnote{Note that  in this section we use `lattice' units, $a_\tau\equiv 1$.} 
\be
    f_s(\tau; \theta_s) = A_s  \cosh\left[ M_s  \left(\tau - \frac{N_\tau}{2} \right) \right],
    \qquad\qquad
    \theta_s=\{M_s, A_s\}.
    \label{eq:1state_model}
\ee
Here $M_s$ represents the mass of the state and $A_s$ its amplitude; these two parameters are collectively referred to as $\theta_s$. Note that the denominator in the single-state propagator (\ref{eq:peak}) is absorbed in the amplitude. 
The population correlation function is then modelled by a combination of several one-state models,
\begin{equation}
\label{eq:CF}
    C(\tau) = \sum_{s = 0}^{N_s - 1} f_s(\tau; \theta_s) \equiv F_{N_s}(\tau; \theta),
\end{equation}
where $\theta=\{\theta_s|s=0,\ldots, N_s-1\}$ is shorthand for the $N_\theta=2N_s$ parameters.
We acknowledge that this model is only strictly valid when the states have a negligible width and no other spectral structures are present. In the analysis we adopt a conservative approach in which its validity is assumed at all temperatures in the hadronic gas. This assumption will be verified a posteriori. Without loss of generality we assume that $M_s<M_{s+1}$, such that $M_0$ is the groundstate mass. 
In our analysis, we fix the number of states included in Eq.~(\ref{eq:CF}) to $N_s = 4$, but we emphasise we are interested in $M_0$ only.

A lattice correlator, denoted as $\hat{C}(\tau)$ in this section, serves as a statistical estimate of
$C(\tau)$. However, as $\hat{C}(\tau)$ is prone to be affected by statistical errors, we model it
using
\begin{equation}
    \hat{C}(\tau) = C(\tau) + u(\tau) = 
        F_{N_s}(\tau; \theta) + u(\tau).
    \label{eq:model}
\end{equation}
Discrepancies between the model and the data are explained by the additive error term $u$. 
 As $\hat{C}(\tau)$ converges in probability to the population correlation function, the expected value
of the noise is zero. It is assumed to be normally distributed but not homoskedastic: the standard deviation of the
noise depends on $\tau$ as the standard error of $\hat{C}(\tau)$ does, with $\mbox{Var}[u(\tau)] = \mbox{Var}[\hat C(\tau)]$ for each $\tau$.

\begin{figure}[t]
      \centering
    \includegraphics[width=0.65\textwidth]{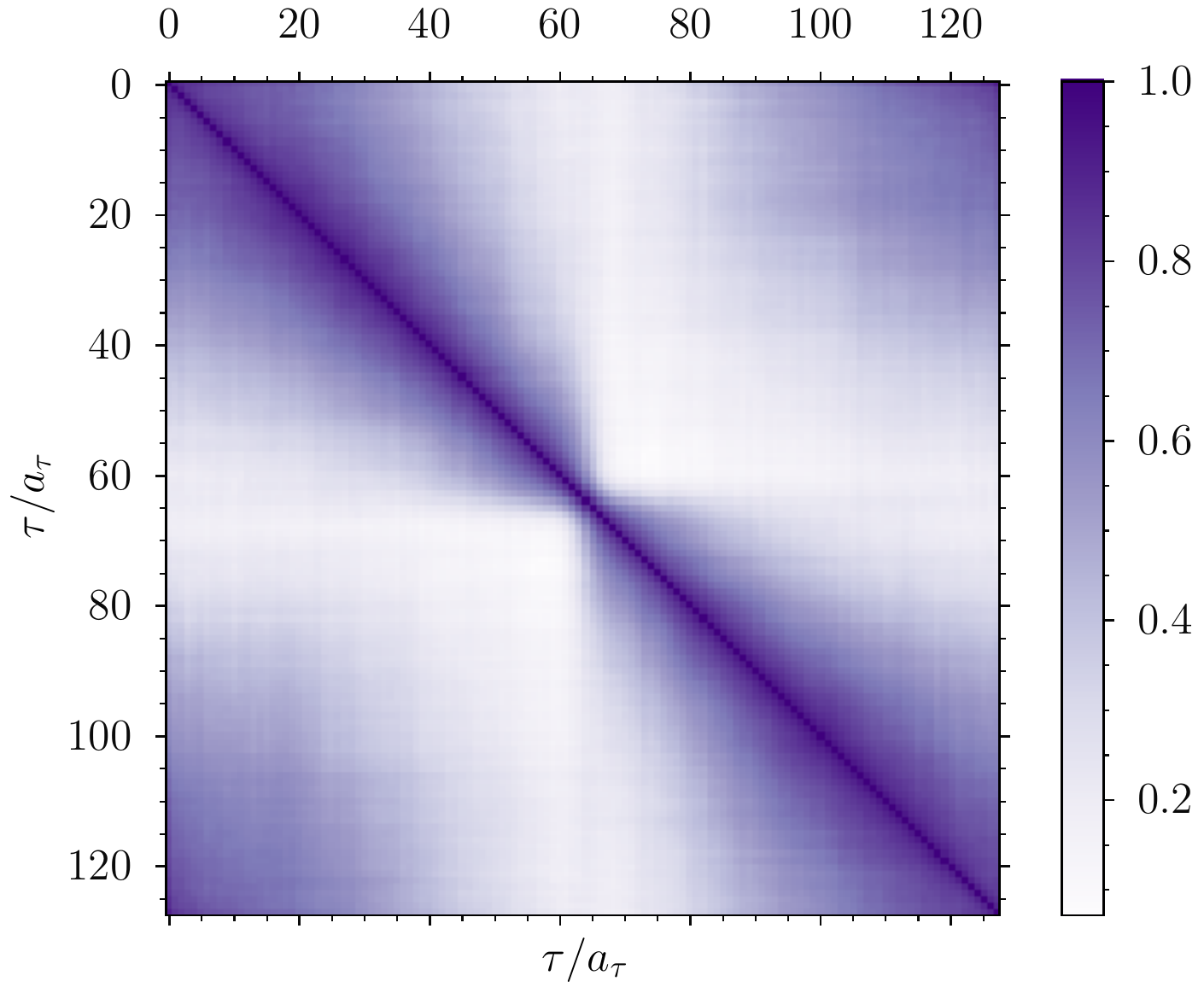}
  \caption{
        Correlation matrix of the $D_s^*$ correlator on the $N_\tau=128$ lattice, computed using Eq.~(\ref{eq:mesons/correlation}).
    }\label{fig:mesons/correlation}
\end{figure}

From now on, we assume that our sample of estimates of $C(\tau)$ are generated by a thermalised
Monte-Carlo chain. Moreover, we also assume that the elapsed computer time between sequential
configuration measurements is long enough so that the Markovian autocorrelation within ensembles is
small. Both assumptions imply that, at fixed $\tau$, all estimates of $C(\tau)$ can be treated as
independent and identically distributed random variables. However, since the same ensemble is
employed to measure all Euclidean times in a particular estimate of $C(\tau)$, the data is expected
to be highly correlated in $\tau$.  This can be demonstrated using the correlation matrix,
\begin{equation}
    X_{ij} = \frac{\Sigma_{ij}}{\sigma_{i} \sigma_{j}},
    \label{eq:mesons/correlation}
\end{equation}
where $\Sigma$ represents the covariance matrix and $\sigma_{i}$ the standard deviation of the
$i^{\rm th}$ estimate in the signal. 
The correlation matrix has a value of $+1$ for linearly correlated variables, and $-1$ for linearly anticorrelated variables.
Indeed, in Fig.~\ref{fig:mesons/correlation} an example of the correlation encountered in lattice data (for the $D_s^*$ correlator in this case) can be seen.

\begin{figure}[t]
       \centering
    \begin{subfigure}[b]{0.32\textwidth}
        \centering
        \includegraphics[width=\textwidth]{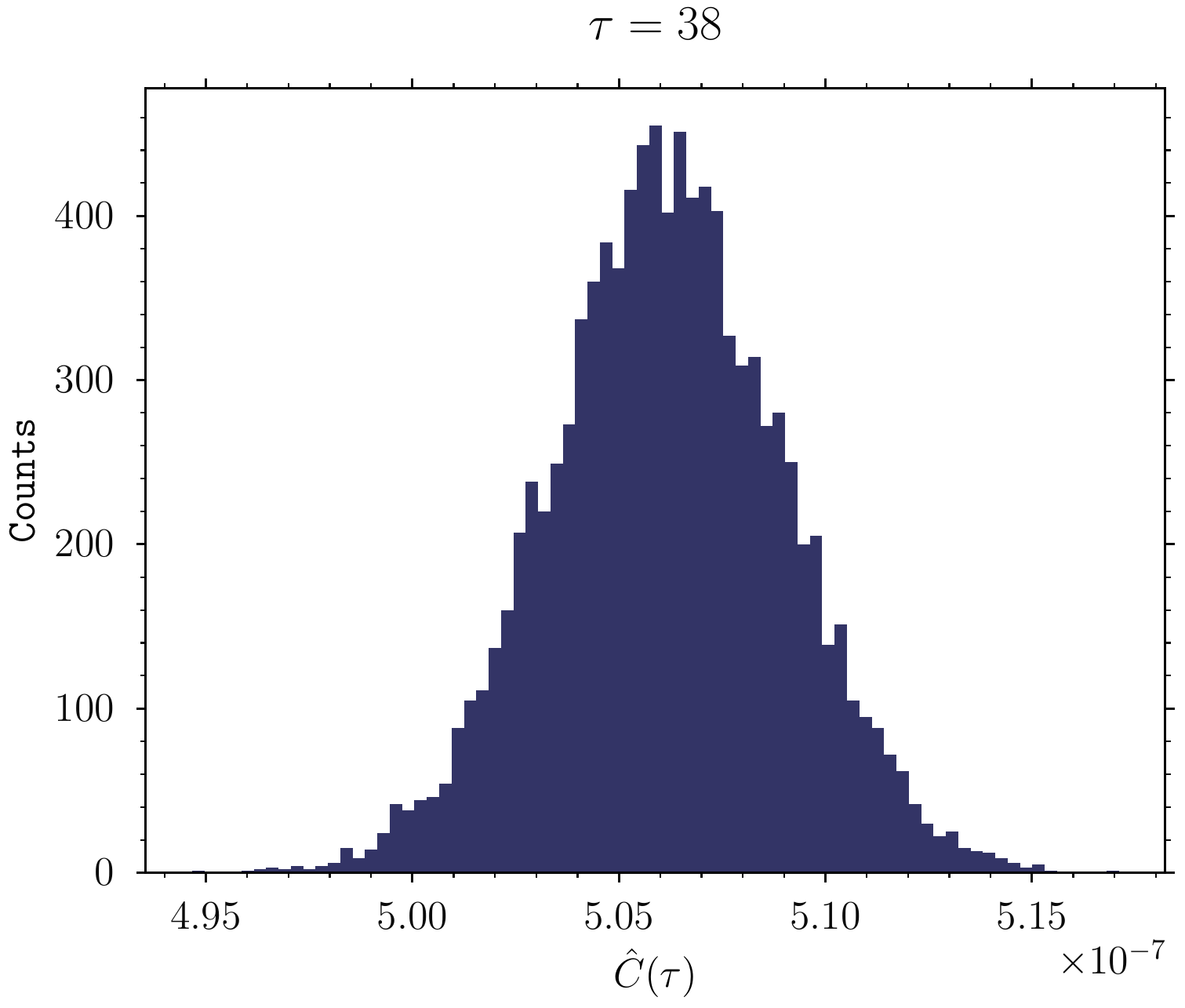}
    \end{subfigure}
    \hfill
  \begin{subfigure}[b]{0.32\textwidth}
        \centering
        \includegraphics[width=\textwidth]{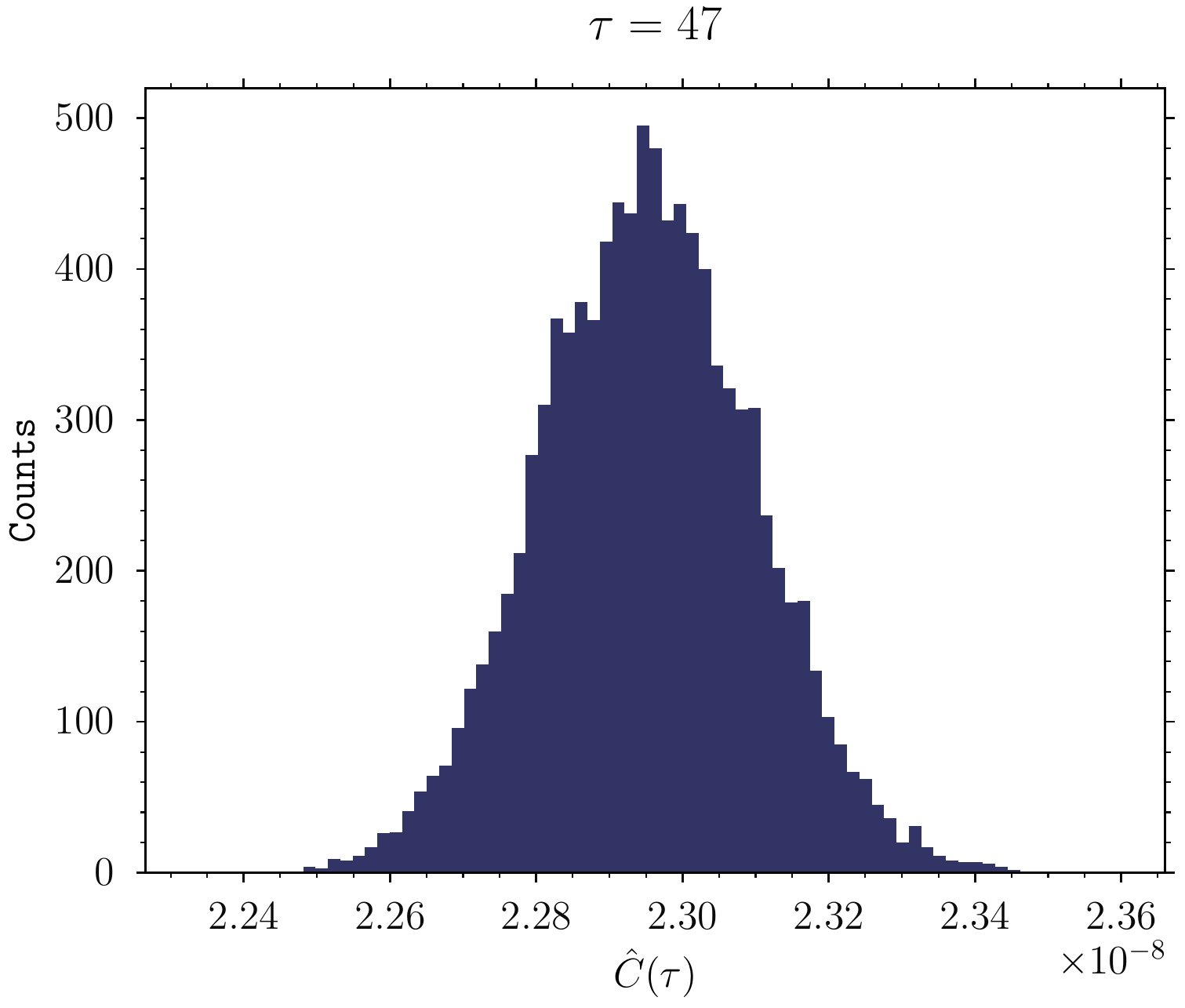}
    \end{subfigure}
    \hfill
    \begin{subfigure}[b]{0.32\textwidth}
        \centering
        \includegraphics[width=\textwidth]{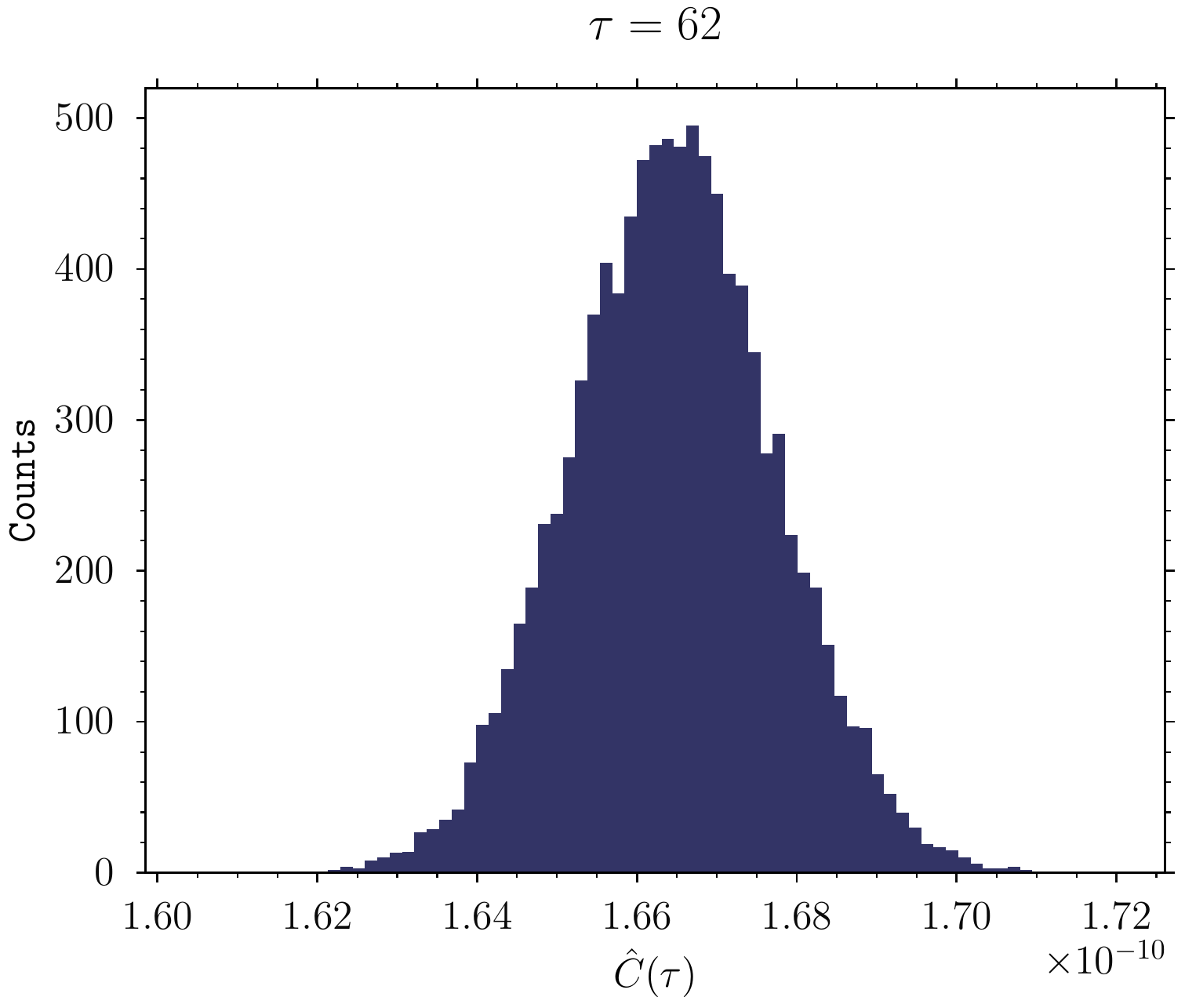}
    \end{subfigure}
       \caption{
        Empirical distribution of the central value of the $D_s^*$ correlator, $\hat{C}(\tau)$, at three
      Euclidean times, $\tau/a_\tau=38, 47, 62$, on the $N_\tau=128$ lattice. The empirical distribution is estimated using
        bootstrap~\cite{efron}.    
         }\label{fig:mesons/distribution}
\end{figure}

As the number of measured ensembles is large enough, and the samples are assumed independent and
identically distributed at fixed $\tau$, we can apply the central limit theorem,
which implies that the sample average is expected to be normally distributed at each independent
$\tau$. Fig.~\ref{fig:mesons/distribution} shows the empirical distribution of $\hat{C}(\tau)$
at three randomly selected Euclidean times for the $D_s^*$ correlator on the $N_\tau=128$ lattice.
As $\hat{C}(\tau)$ is normally distributed, at fixed $\tau$, $u$ is also normally distributed, with 
mean $\vec{0}$  and covariance $\Sigma$; the covariance of $u$ and $\hat{C}(\tau)$ is equal. The total 
likelihood of the data can then be modelled using a multivariate correlated normal distribution,
\begin{equation}
    P(u | \theta) = 
        \mathcal{N}(\mu = 0, \Sigma) = 
        \mathcal{N}(\hat{C}(\tau) - F_{N_s}(\tau; \theta), \Sigma).
\end{equation}
In the equation above, $\mu = \hat{C}(\tau) - F_{N_s}(\tau; \theta) = \vec{0}$ has dimensions $N_\tau$,
while $\Sigma$ has dimensions $(N_\tau \times N_\tau)$.

Finding the values of the parameters that maximise this likelihood function is equivalent to finding 
the set of parameters $\hat{\theta}$ that minimises the following correlated target scalar function,
\begin{equation}
    \mathcal{L}(\theta) =  \sum_{\tau,\tau'} \left[\hat{C}(\tau) - F_{N_s}(\tau, \theta)\right]
                \Sigma(\tau, \tau^{\prime})^{-1}
                \left[\hat{C}(\tau^\prime) - F_{N_s}(\tau^\prime, \theta)\right].
    \label{eq:loss}
\end{equation}
Due to the inherent complexity of Eq.~(\ref{eq:loss}), the minimisation must be performed
numerically.

Once the maximum likelihood parameters are estimated, their uncertainties can be approximated using
the so-called Fisher information function~\cite{lehmann2006theory}. In our particular case, we can
approximate the covariance of $\hat{\theta}$ using
\begin{equation}
    \text{Cov}(\theta_a, \theta_b) =  \sum_{\tau,\tau'} 
        J(\theta_a, \tau) \Sigma(\tau, \tau^{\prime}) J(\theta_b, {\tau^\prime}),
\end{equation}
where $J$ represents the Jacobian of $\hat C(\tau)$ with respect to $\theta$, which can be computed using
the first derivatives,
\begin{equation}
    J(\theta_a, \tau) = \frac{\partial \hat{C}(\tau)}{\partial \theta_a}
                      = \frac{\partial F_{N_s}(\tau, \theta)}{\partial \theta_a}.
\end{equation}

Finally, to improve the stability of the regression, we fold the correlation function estimates
around the midpoint in the temporal direction, 
\begin{equation}
    \hat{C}_F(\tau) = \half\left[ \hat{C}(\tau) + \hat{C}(N_\tau - \tau)\right].
\end{equation}
This can be viewed as a data augmentation procedure, which is justified by the symmetry of the
mesonic propagators under Euclidean time inversion. 
In the analysis, we also normalise our lattice estimates by $\hat{C}(\tau = N_\tau / 2)$.

\subsection{Initial parameter estimation}

Correlated fits tend to be unstable: slight variations in the initial values of the parameters may
lead to different local minima of Eq.~(\ref{eq:loss}). It is hence important to provide our
minimisation routine with a good set of initial parameters close to their underlying population
values. To obtain these parameters, we perform a combination of effective mass computations and fits
to models with different number of states. 
The models are sub-models of the $N_s$-state model in
Eq.~(\ref{eq:model}), obtained by truncating Eq.~(\ref{eq:model}) at  $N_s^{\rm sub} \leq N_s$, 
while the effective mass can be extracted by solving the following transcendental equation,
\be
    \frac{\hat{C}(\tau+1)}{\hat{C}(\tau)} 
        = \frac{\cosh\left[ M^{\rm eff}  \left(\tau+1 - N_\tau/2\right)\right]}
        {\cosh\left[M^{\rm eff}  \left(\tau - N_\tau/2 \right)\right]}.
    \label{eq:mesons/effmass}
\ee
The mass of the $s^{\rm th}$ state can be estimated by computing the
effective mass of the correlation function if knowledge about the parameters of the $s -1$ lighter
states is available, such that a subtracted correlator can be constructed, 
\be
    \hat C_{s}(\tau) = \hat{C}(\tau) - F_{s}(\tau; \theta).
\ee
The groundstate of $\hat C_s(\tau)$ corresponds to the mass of the $s^{\rm th}$ state as long as the
previously estimated parameters are correctly determined. As uncertainties propagate, this procedure
becomes unreliable for heavy states. As a result, we heuristically initialise higher-order states
masses using
\be
    M_s = 1.5  M_{s - 1} \qquad\qquad (s>2).
\ee
Once an estimate of the mass of the $s^{\rm th}$ state is determined, we produce a series of
correlated fits to $\hat{C}(\tau)$ as a way of improving our estimate mass estimate. Additionally,
these fits allow us to produce an estimate of amplitudes of each state, which are always initialised
with a value of $1$.

\subsection{Regression at a particular fit window}

With the initial parameters estimated, we proceed with the multi-model regression analysis.  We use a large collection of fit windows, defined by the intervals $\tau_0 \leq \tau \leq \tau_f$, or
\be
    FW[\tau_0, \tau_f] = [\tau_0, \tau_f].
\ee
For the starting point we use $2-5 \leq \tau_0 \leq \tau_f - 3$, with the lower bound depending on the correlation function and temperature; 
the final timeslice $\tau_f$ in the fit window is usually taken as $\tau_f = N_\tau/2$ as a way of maximising the
amount of information included in the regression. However, when the signal gets noisy (typically at large $\tau$ at large $N_\tau$)  it is beneficial to also vary $\tau_f$. This is further discussed in the next subsection.

Here we describe how a groundstate mass is extracted at a particular fit window $FW[\tau_0, \tau_f]$. 
Independent estimates of the groundstate mass are produced employing different sub-models with $N_s^{\rm sub}\leq N_s$ states.
The models are then fitted to the same data using the correlated target function defined in
Eq.~(\ref{eq:loss}). However, not all models are included in all fit windows. For a given fit window, we only include
models that fulfil the following two conditions: 
\begin{itemize}
\item
the size of the fit window,  $\tau_f - \tau_0+1$, is larger than the number $2N_s^{\rm sub}$ of free parameters in the model;
\item 
it is assumed realistic enough to model contributions of higher-order states if they are expected in the fit window.
\end{itemize}
The first condition ensures that we are fitting a given model to enough information, while the second one ensures that oversimplified models are not used in cases where higher-order states contribute. 

For each model included in the regression, we obtain an estimate of the groundstate mass. In order
to generate a model-independent groundstate mass at a given fit window, we compare all the
groundstate masses using the so-called {\em corrected Akaike Information Criterion} ($\AICc$) \cite{akaike1974new, akaike1998information, anderson2004model}.
The $\AICc$ has its roots in information theory, and measures the expected divergence between a
model and an unknown ground truth model; the divergence is measured using the Kullback-Leibler (KL)
divergence~\cite{kullback1951information}. In our definition of the $\AICc$, the model with the
lowest $\AICc$ among all is the most likely to correctly describe the data. It is defined by 
\begin{equation}
    \AICc = 
        N_\theta - \log(\hat{L}) + \frac{N_\theta^2 + N_\theta}{(\tau_f-\tau_0+1) - (N_\theta+1)},
\end{equation}
where $N_\theta$ corresponds to the number of free parameters of the model; $\hat{L}$ is the
objective function defined in Eq.~(\ref{eq:loss}), and evaluated at the maximum likelihood
parameter estimates; and $\tau_f-\tau_0+1$ represents the number of points included in
the current fit window. Note that the $\AICc$ metric depends on the chosen model through $N_\theta$
and $\hat{L}$.

The $\AICc$ can be used to compute the relative model quality between models $F^\prime$ and $F$,
\begin{equation}
    l(F^\prime, F) = \exp\left(-\frac{1}{2} \big[\AICc(F^\prime) - \AICc(F)\big]\right).
\end{equation}
This quantity measures how likely $F'$ is to correctly model the data when compared to $F$. Note that $l$ does not include global information. As a result, there exists the possibility that models not included in the analysis might be better at describing the data at the given fit window. 

Provided that we fix $F$ to be model with the lowest $\AICc$, then the relative model quality can be
employed to weigh the quality of each groundstate mass estimation. As a result, we can compute a
final estimate of the groundstate mass at the given fit window by computing a weighted average
between all groundstate estimates, using $l(F^\prime, F)$ as the weight. This technique allows us
to promote the influence of high-quality models in our final result while, at the same time,
avoiding manually discarding any models. Finally, the error in the weighted mass can be approximated
using a bootstrap analysis.

The result of this analysis is an estimate of the groundstate mass at a given fit window, which
we label as $\hat{M}_0[\tau_0, \tau_f]$.

\subsection{Varying the final timeslice}

\begin{figure}[t]
       \centering
    \begin{subfigure}[b]{0.32\textwidth}
        \centering
        \includegraphics[width=\textwidth]{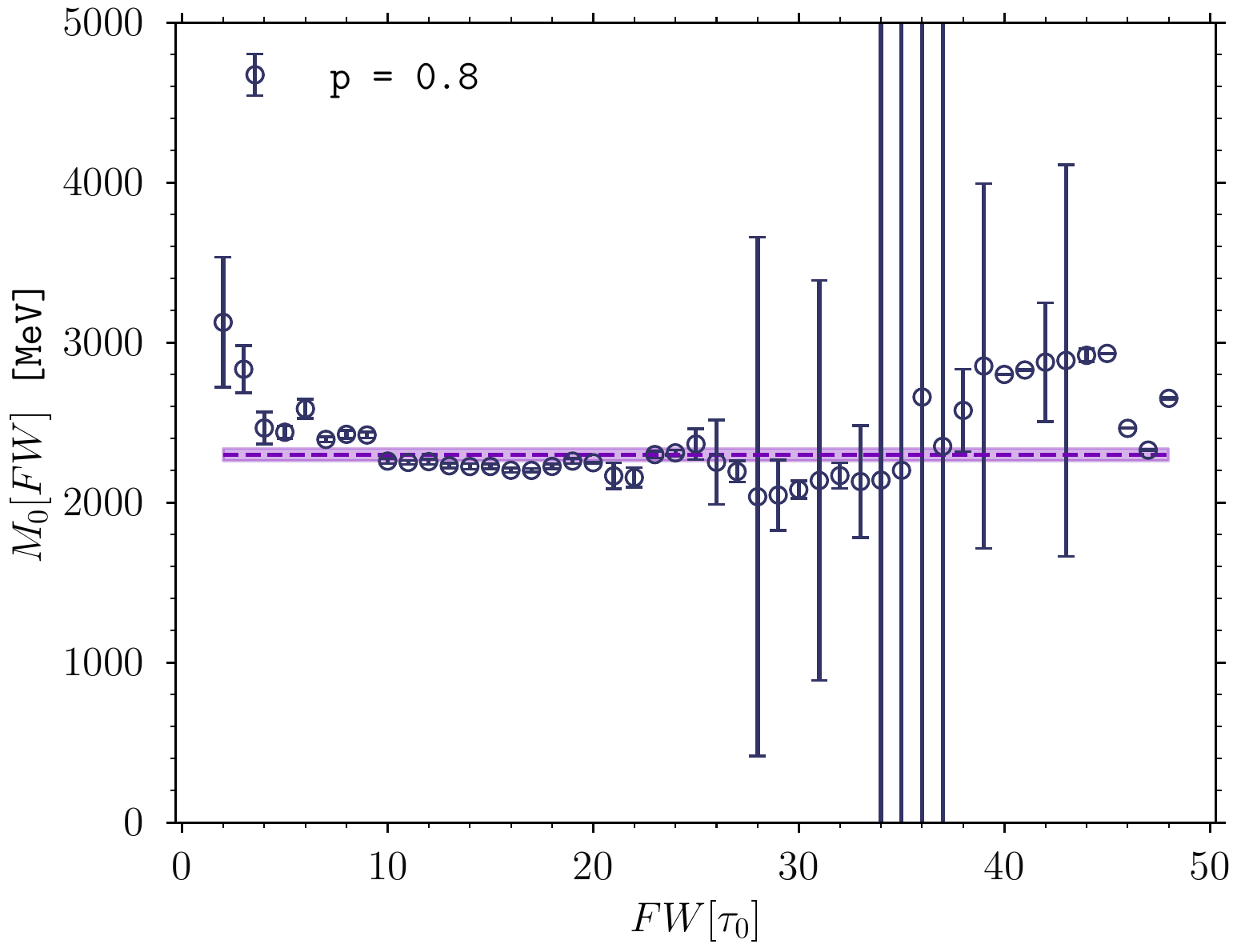}
        \caption{$p = 0.8$}
    \end{subfigure}
    \hfill
    \begin{subfigure}[b]{0.32\textwidth}
        \centering
        \includegraphics[width=\textwidth]{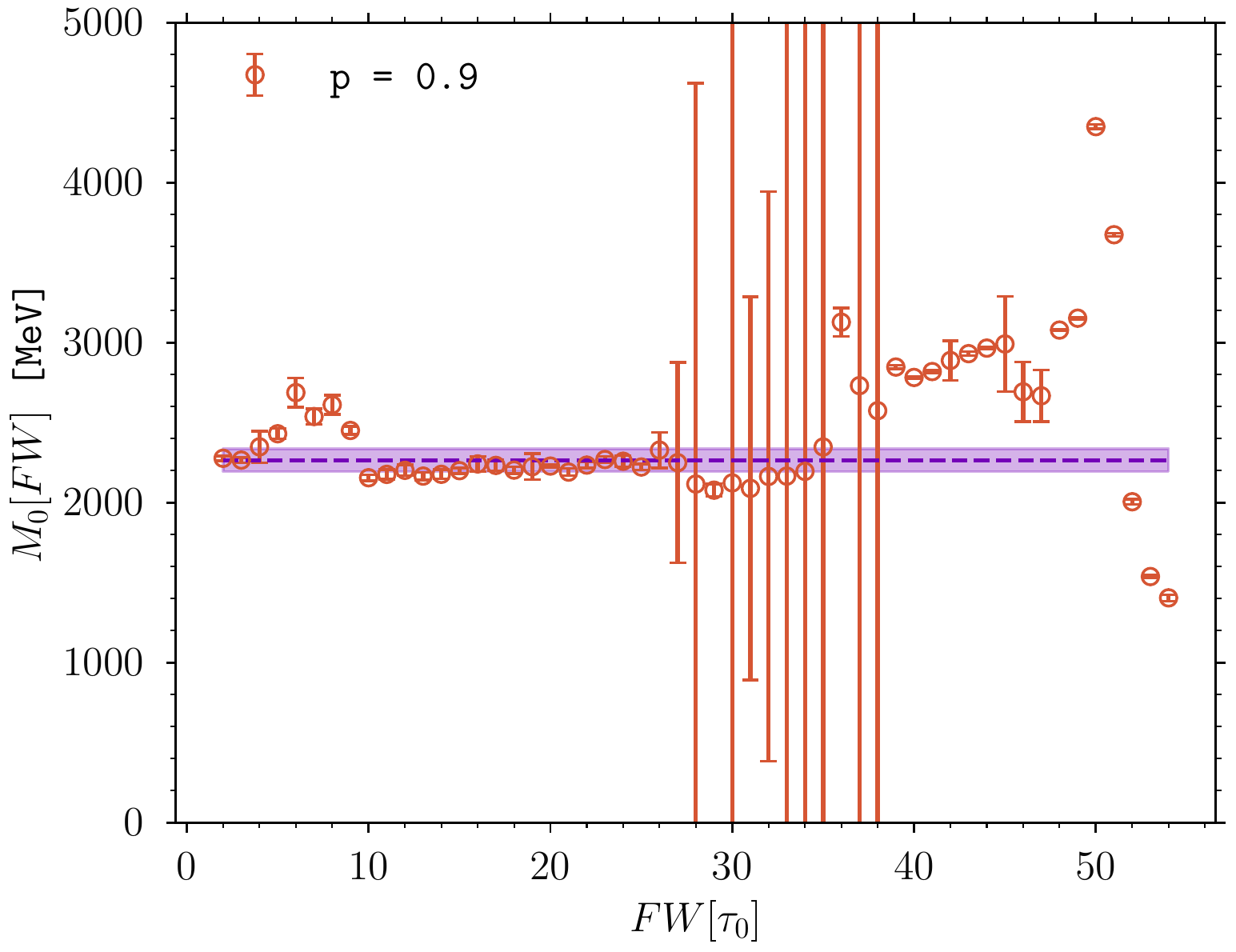}
        \caption{$p = 0.9$}
    \end{subfigure}
    \hfill
    \begin{subfigure}[b]{0.32\textwidth}
        \centering
        \includegraphics[width=\textwidth]{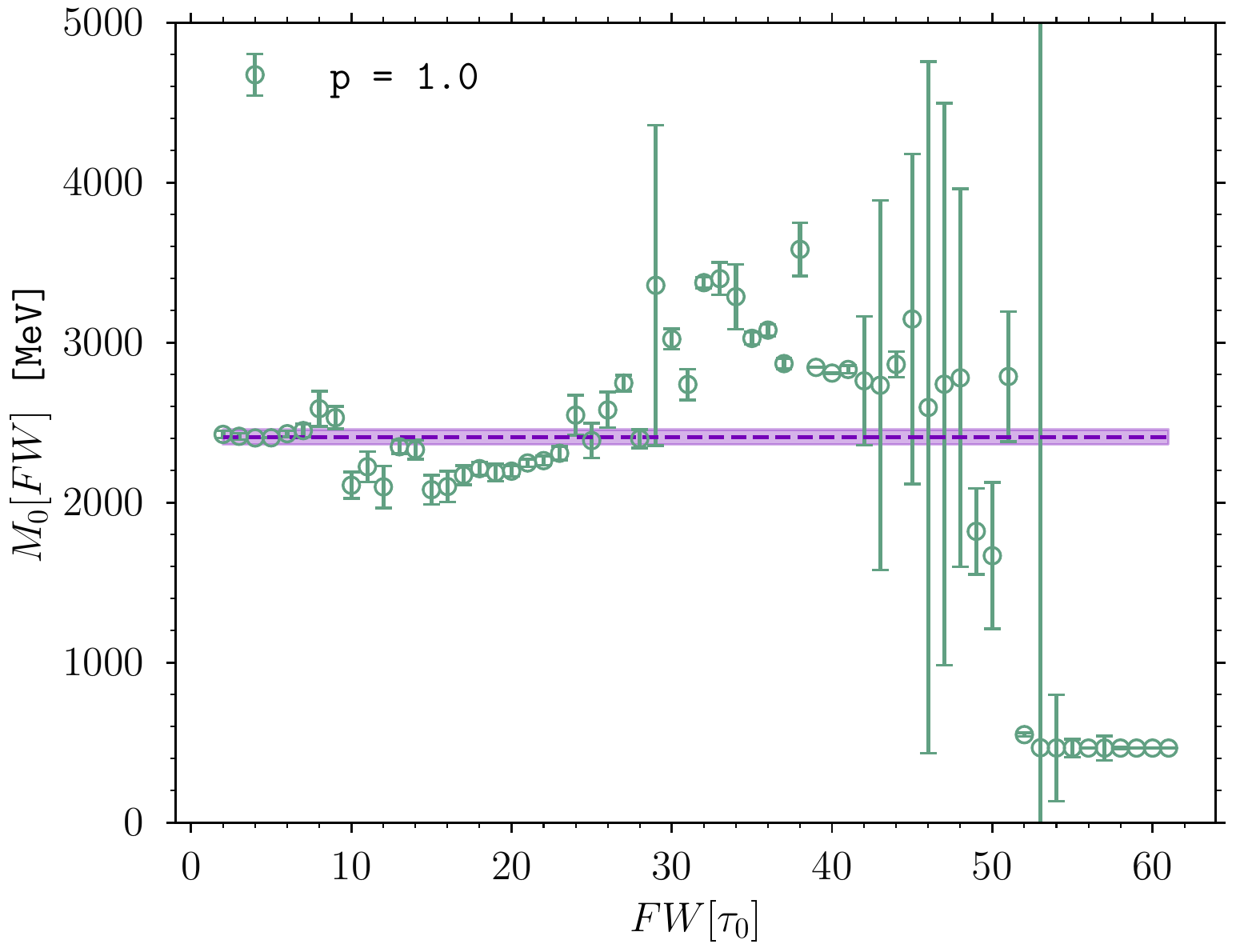}
        \caption{$p = 1.0$}
    \end{subfigure}
    \\
    \includegraphics[width=0.50\textwidth]{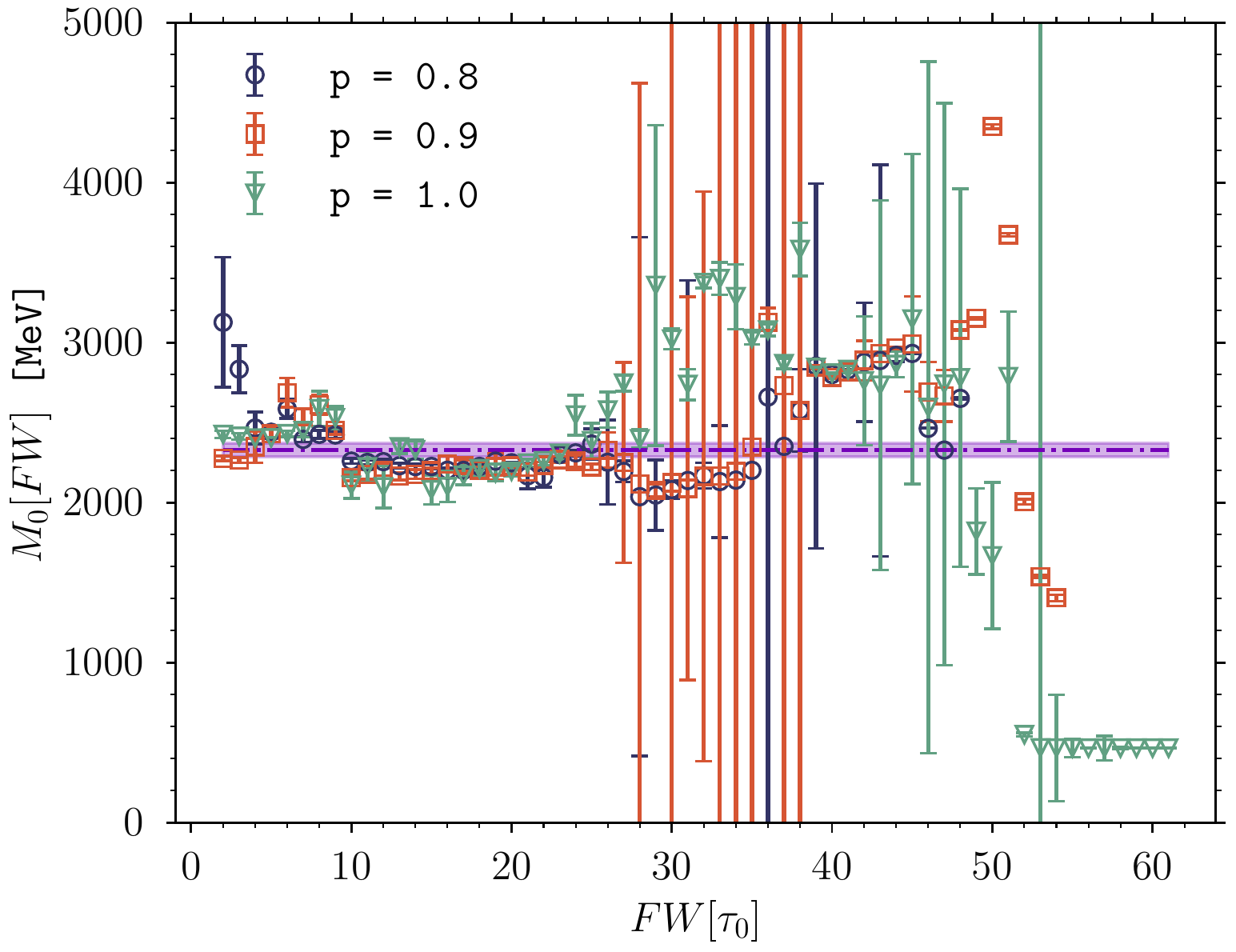}
 \caption{
        Analysis performed for the $D_1$ correlation function on the $N_\tau=128$ lattice.
        The first row shows the groundstate masses extracted at fixed $\tau_f=51, 57, 64$ ($p=0.8, 0.9, 1$). The second row
        contains the combined estimate.
    }\label{fig:mesons/pstudy}
\end{figure}

It is preferable to take as final timeslice in the fit window the largest possible value, i.e.\ $\tau_f=N_\tau/2$, to both optimise the number of points as well as the overlap with the groundstate. In some channels however, the data is noisy, especially on large lattices ($N_\tau=128$). We therefore have also varied $\tau_f$, using 
\be
    \tau_f = p  \frac{N_\tau}{2}, 
    \qquad\qquad p = 0.8, 0.9, 1.0. 
    \label{eq:p_value}
\ee
By selecting different values of $\tau_f$, we can avoid noisy points in the analysis. However, one
of our premises is to generate a methodology that avoids manually selecting any results. 
We use therefore the following procedure to compute a systematic
$\tau_f$-independent groundstate mass: first, we independently apply the previously discussed
methodology using different fixed values of $\tau_f$, which generates a set of groundstate
estimates depending on $\tau_f$, $\hat{M}_0[\tau_0, \tau_f]$; then, we combine all the estimates into a
single sample; to conclude, we compute the median estimate of the combined sample after removing the
outliers outside the IQR interval. No fit window is included more than once in the sample.
Fig.~\ref{fig:mesons/pstudy} shows an example of this procedure in the case of the $D_1$ groundstate on the $N_\tau=128$ lattice.

The median of the combined sample represents the best estimate of the groundstate mass that we can
produce. This procedure can be applied as we are always extracting an estimate of the same
underlying population groundstate mass, $M_0$, independently of $\tau_f$; i.e.\ the same model is valid at all Euclidean times.

\subsection{Final groundstate mass}

After applying the algorithm described above for each fit window available, we are left with a
collection of estimates of the same underlying groundstate mass. This collection contains all the
results extracted at all fit windows generated by varying $\tau_0$ with different fixed finishing
times $\tau_f$, computed using Eq.~(\ref{eq:p_value}). In practice, we have access to the following
collection,
\begin{equation}
    \mathcal{M}_0 = \left\{ 
        \hat{M}_0[\tau_0, \tau_f]\, | \, 
            \tau_0 \in [2-5, \tau_f - 3],\, 
            \tau_f \in (0.8, 0.9, 1.0)  N_\tau / 2
        \right\}.
\end{equation}
No fit window is included twice in $\mathcal{M}_0$.

As we assume that Eq.~(\ref{eq:model}) is the true model for all $\tau$, then the groundstate
masses extracted at all fit windows correspond to an independent estimate of the same underlying
population parameter, $M_0$. As a result, we can combine all the estimates into a single sample. The
median of this sample represents the best estimate of the groundstate mass that we can obtain with
our procedure. We compute the median as outliers tend to be present in $\mathcal{M}_0$ due to the
unstable nature of correlated fits and the lack of prior information in the estimation of initial
parameters. Although the median is a robust statistic, we discard the outliers in our sample, which
are spotted using a standard Interquartile Range (IQR) condition. The uncertainty of our final
groundstate mass can be approximated using bootstrap.

This procedure yields the best estimate of the groundstate mass in lattice units, i.e.\ $a_\tau \hat M_0$. To convert to physical units, we use that the inverse lattice spacing equals $a_\tau^{-1}=6079(13)$ MeV, and combine the uncertainties from the regression analysis and the scale setting in the standard way, adding relative uncertainties in quadrature.

\end{document}